\documentclass{jfm}
\usepackage{graphicx}
\usepackage{epstopdf, epsfig}
\usepackage{upgreek} 
\usepackage{bm} 
\usepackage{mathtools} 
\usepackage{mathrsfs}
\usepackage{color}
\usepackage{xcolor}
\usepackage{tabularx}

\newcommand\putfig[2]{\begin{tabular}[t]{@{}l@{}}#1\\#2\end{tabular}}

\shorttitle{Modulation of turbulence regeneration cycle by inertial pointwise particles}
\shortauthor{G. Wang and D. H. Richter}

\title{Modulation of the turbulence regeneration cycle by inertial particles in planar Couette flow}
\author{
  G. Wang
  \corresp{\email{gwang4@nd.edu}}
 \and 
  D. H. Richter
  \corresp{\email{David.Richter.26@nd.edu}}  
 }

\affiliation{Department of Civil and Environmental Engineering and Earth Sciences, University of Notre Dame, Notre Dame, Indiana 46556, USA}

\begin{document}

\maketitle

\begin{abstract}

Two-way coupled direct numerical simulations are used to investigate the effects of inertial particles on self-sustained, turbulent coherent structures (i.e. the so-called the regeneration cycle) in plane Couette flow at low Reynolds number just above the onset of transition. Tests show two limiting behaviors with increasing particle inertia, similar to the results from the linear stability analysis of \citet{saffman1962stability}: low-inertia particles trigger the laminar-to-turbulent instability whereas high-inertia particles tend to stabilize turbulence due to the extra dissipation induced by particle-fluid coupling. Furthermore, it is found that the streamwise coupling between phases is the dominant factor in damping the turbulence and is highly related to the spatial distribution of the particles. The presence of particles in different turbulent coherent structures (large scale vortices or large scale streaks) determines the turbulent kinetic energy of particulate phase, which is related to the particle response time scaled by the turnover time of large scale vortices. By quantitatively investigating the periodic character of the whole regeneration cycle and the phase difference between linked sub-steps, we show that the presence of inertial particles does not alter the periodic nature of the cycle or the relative length of each of the sub-steps. Instead, high-inertia particles greatly weaken the large-scale vortices as well as the streamwise vorticity stretching and lift-up effects, thereby suppressing the fluctuating amplitude of the large scale streaks. The primary influence of low-inertia particles, however, is to strengthen the large scale vortices, which fosters the cycle and ultimately reduces the critical Reynolds number. \\

\end{abstract}


\section{Introduction}\label{sec:Introduction}

Turbulence in the atmospheric boundary layer (ABL) is responsible for the dispersion of pollutants, dust, sand and other constituents \citep{garratt1994atmospheric}, however even with dilute volume concentrations, turbulence modulation by particles still poses a formidable challenge to fully understanding how these constituents alter their own transport \citep{balachandar2010turbulent}. 
Turbulence in the ABL is characterized by very high Reynolds numbers, and is therefore subject to the numerous interactions between inner and outer layers, including the dynamics of the so-called very large scale motions (VLSMs) which develop \citep{smits2011high,inoue2012inner,jimenez2018coherent}. Since in many atmospheric multiphase flows the dispersed phase is emitted from the surface (e.g. sand saltation, bubbles bursting at the water surface), the effect of particulates on inner-outer interactions and more specifically on the regeneration cycles of near-wall coherent motions is the primary interest of the present study.
While the complexity of high-Reynolds-number, multiphase, turbulent flows is high due in large part to the wide range of relevant spatial and temporal scales, the present aim is to simplify the problem and examine one key aspect of the larger picture: the regeneration cycle of coherent structures in the overlap region of wall-bounded turbulence. In this regard, turbulent plane Couette flow at low Reynolds number is known to contain the self-sustained processes exhibited by turbulent coherent structures \citep{waleffe1997self, hamilton1995regeneration}, as well as large-scale structures interacting with smaller near-wall motions \citep{komminaho1996very, kitoh2005experimental}. This setup is therefore a representative, computationally inexpensive candidate for shedding light on particle transport and the fundamental modification of the regeneration cycle and inner/outer interactions. 


\begin{figure}
  \centerline{\includegraphics[trim = 0mm 0mm 0mm 0mm,width=12.3cm]{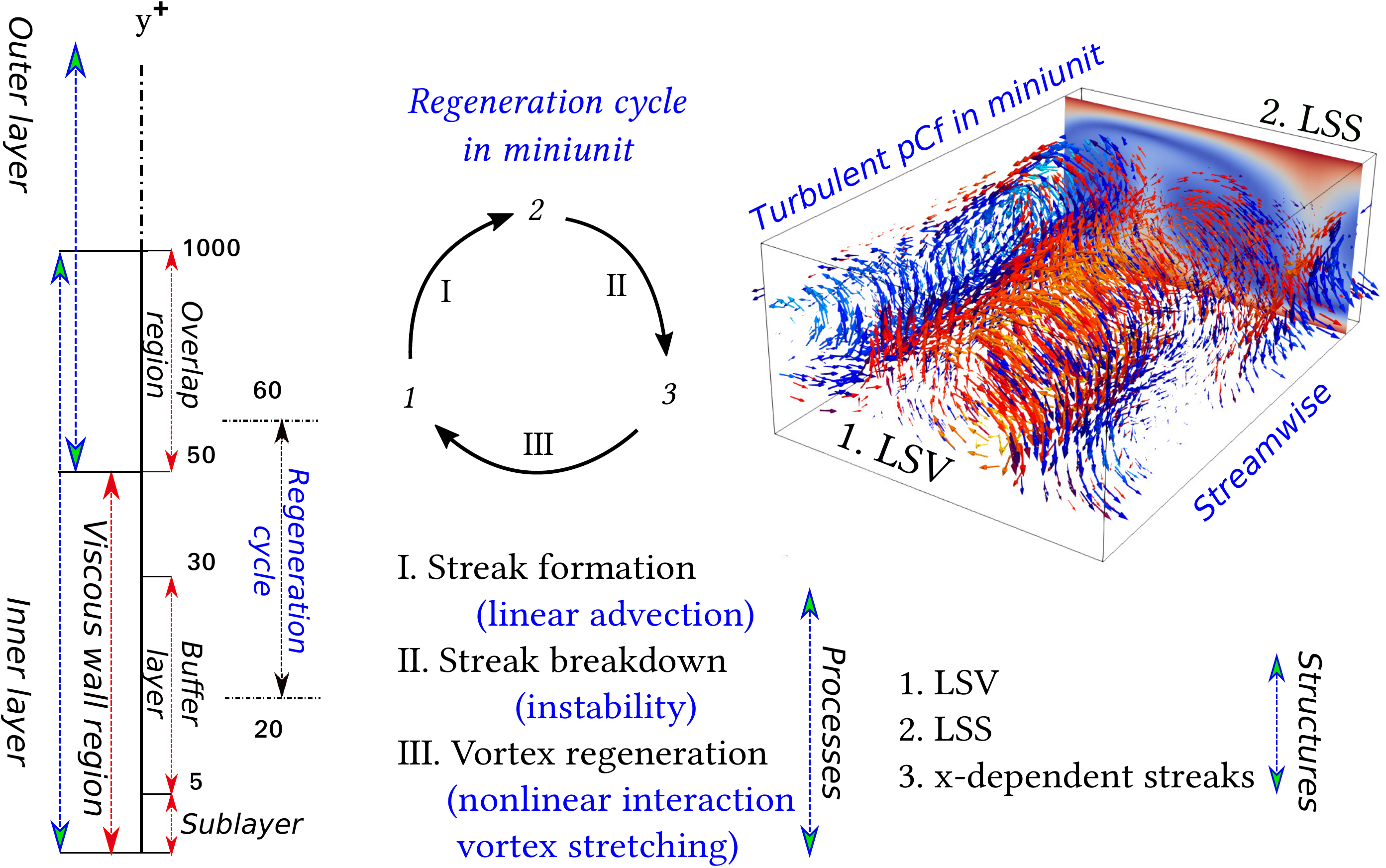}}
  \caption{Sketch of various wall regions and boundary layers in wall units \citep{pope2000turbulent,jimenez1999autonomous} and the regeneration cycle sub-steps \citep{hamilton1995regeneration, waleffe1997self, schoppa2002coherent} in a domain size called `minimal' unit. In the top-right is the flow field in turbulent plane Couette flow in miniunit, large-scale vortex (LSV) is shown by vector fields and colored by the streamwise vorticity. Large-scale streak (LSS) is shown by the contour of the streamwise velocity magnitude.}
\label{fig:Figure_regenerationcycle}
\end{figure}

Currently, it is commonly accepted that the full regeneration cycle in the inner layer exhibits the main characteristics of turbulence \citep[]{hamilton1995regeneration, waleffe1997self, jimenez1999autonomous, kawahara2001periodic, schoppa2002coherent}. Figure \ref{fig:Figure_regenerationcycle} provides a schematic of the various regions near the wall, as well as the key turbulent structures and processes which make up the regeneration cycle in the region of $20<y^+<60$; see \citet{waleffe1997self} and \citet{jimenez1999autonomous}. This self-sustained regeneration cycle is marked by three important structures: large-scale vortices (LSVs), large-scale streaks (LSSs), and x-dependent meandering streaks; each stage of the regeneration process is associated with sequential sub-processes. Specifically, the streaks are generated by a linear process, the so-called lift-up effect, whereas the following two processes are the result of nonlinear interactions called streak breakdown and vortex regeneration. The regeneration cycle was first investigated by \citet{hamilton1995regeneration} in plane Couette flow using a so-called `miniunit' configuration, which is the minimal geometric domain that is sufficient to accommodate the self-sustained flow structures for single-phase turbulence. This minimal simulation domain was carefully examined in Couette flow configuration by \citet{hamilton1995regeneration} and pressure-driven flow by \citet{jimenez1991minimal}. In both cases, the spanwise length is slightly larger than $100$ wall units, which corresponds to the spanwise characteristic spacing between two coherent structures.

The modulation of the regeneration cycle has been found to be related to the spatial distribution of a particulate phase. In fully-resolved simulations with finite-sized, neutrally-buoyant particles, the cycle is promoted in turbulent pressure-driven flow, and is hardly modified in turbulent plane Couette flow \citep{brandt2014lift,wang2018JFM}. In the case of plane Couette flow, particles are more likely to be present in the core of the large scale rolls which are mainly responsible for dissipating energy. The same phenomenon is also observed experimentally in Taylor-Couette flow \citep{majji2018inertial} if the particle concentration is low and and the particles are relatively small compared with the gap. However in pressure-driven flow, finite-size, neutrally-buoyant particles show a non-monotonic effect on laminar-to-turbulent transition. Here, a low volumetric concentration (less than $20\%$) of large particles (particle to pipe diameter ratio around $1/20$) sustains the turbulent state and decreases the transition threshold, whereas a high concentration of small particles advances the transition threshold significantly \citep[see][]{matas2003transition}. \citet{wang2018JFM} further found that particles tend to trigger instability in pressure driven flow due to the particles' preferential presence in near-wall streaks.

In general, turbulence modulation is found to depend on the size ratio between the particle and turbulence length scales, and on the time scale ratio between the particle response time with a characteristic time scale of the flow --- i.e., the dimensionless Stokes number $St$ \citep{gore1989effect,elghobashi1994predicting}. \citet{gore1989effect}, for instance, proposed a relationship of size ratio between particle diameter and the flow integral length scale to flow turbulent intensity attenuation/augmentation. \citet{tanaka2008classification} formulated a dimensionless particle momentum number to predict turbulence modulation which includes the flow microscale (Kolmogorov scale), the energy-containing scale, as well as the Stokes number. At high Reynolds number, \citet{tanaka2008classification} divided the turbulent modification into three regions: moderate momentum number tends to attenuate the turbulence whereas low or high momentum number tend to augment the turbulence. For heavy particles which can be reasonably modeled using only a drag force (i.e. neglecting lift, added mass, etc. \citep{maxey1987motion}), \citet{balachandar2010turbulent} proposed that the turbulence modulation mechanisms can be simplified: turbulence reduction comes from enhanced inertia and increased dissipation arising from particle drag, whereas the turbulence enhancement is due to enhanced velocity fluctuation from wake dynamics.
A large body of work on turbulent kinetic energy modification \citep{squires1990particle, elghobashi1993two, Pan1996PoF, burton2005fully}, interphasial energy transfer \citep{zhao2013interphasial}, particle transfer and segregation \citep{marchioli2002mechanisms}, or drag reduction \citep{li2001numerical, dritselis2008numerical} can be found in the literature.

While we are ultimately interested in the modulation of high-Reynolds number wall turbulence due to inertial particles, our specific focus in this work is on the modification of the near-wall regeneration cycle and its close connection with laminar-to-turbulent transition as a model for inner/outer interactions; this influence of a dispersed phase in turbulent flows has received relatively little attention. At Reynolds numbers close to the onset of transition, \citet{klinkenberg2011modal} proposed a stability Stokes number, where at small values (fine particles) the critical Reynolds number decreases proportionally to the particle mass fraction, and where at intermediate values yields an increase of the critical Reynolds number where the enhancement is proportional to the volume fraction. Through linear analysis of the Orr-Sommerfeld equation coupled to a dispersed phase via drag, \citet{saffman1962stability} theoretically predicted that fine particles tend to destabilize the gas flow whereas coarse particles have a stabilizing action. The drag coupling between phases therefore has competing effects: low inertia particles tend to destabilize the flow by adding to the effective density of the gas, and at high inertia particles tend to act as an extra dissipation source, thereby stabilizing the flow.
Following the formulation proposed by \citet{saffman1962stability}, \citet{michael1964stability}, \citet{rudyak1997hydrodynamic}, and \citet{klinkenberg2011modal} numerically solved the modified Orr-Sommerfeld equation for plane Poiseuille flow, and \citet{michael1964stability} showed an increased critical Reynolds number at low particle concentration whereas \citet{rudyak1997hydrodynamic} and \citet{klinkenberg2011modal} found that the critical Reynolds number reaches a maximum and then decreases gradually with increasing particle relaxation time.
Other configurations have also been investigated. \citet{dimas1998linear} studied the particle laden mixing layer and \citet{despirito2001linear} analyzed a particle-laden jet. A similar conclusion is drawn that small, low-inertia particles induce a destabilization effect whereas large Stokes number particles stabilize, and that the effect is approximately proportional with the particle mass loading (at least in dilute concentrations). For these stability analyses, however, the particles often must be assumed to be homogeneously distributed or distributed with an assumed profile; nonlinear effects such as particle preferential accumulation and the dynamics of the near-wall regeneration cycles are difficult to include in the analysis. Some nonlinear studies on transition do exist, including those of 
\cite{klinkenberg2011modal,klinkenberg2013numerical}, who also find that small inertia tends to decrease the critical Reynolds number whereas particles with intermediate Stokes numbers increase the critical Reynolds number. From dilute limit to high-mass-loading, \citet{capecelatro2018transition} find that fluid-phase turbulence kinetic energy is generated by mean-shear production in dilute limit ($\overline{\Phi_m} \leq 1$), whereas it is entirely generated by drag production at high mass loading ($\overline{\Phi_m} \geq 10$). In the intermediate regime ($2 \leq \overline{\Phi_m} \leq 4$), the flow relaminarizes due to higher rate of dissipation compared to production of turbulence kinetic energy. Mechanisms regarding the disruption of turbulence regeneration, however, are still an open question. 

To study the particle-induced modulation of the self-sustained regeneration cycle in wall turbulence, numerical simulations are performed for particle suspensions in plane Couette flow in a `miniunit' domain. 
In previous work, finite-size particles with low Stokes number are found to hardly modify the regeneration cycle due to particles preferentially accumulating in the large-scale vortices \citep{wang2018JFM}. At a higher Reynolds number, however, pointwise inertial particles with moderate to high Stokes numbers preferentially `operate' on flow structure scales associated with the particle response time scale \citep{richter2013momentum, richter2014modification,richter2015turbulence}, but this effect on the regeneration cycle remains unclear.
This paper is organized as follows. Section \ref{sec:Method} presents the numerical method used in this work and validation for single-phase flow in the miniunit. Section \ref{sec:Configuration} summarizes the parameter choices and in Section \ref{sec:Statistics}, we carefully perform a set of numerical tests to observe the inertial effects on transition by varying particle response time, mass fraction, and volume fraction. Some important statistical quantities are further computed and analyzed in this section. We then focus on the particle effects on the regeneration cycle during turbulence enhancement/attenuation through modal analysis in Section \ref{sec:regeneration_cycle}.\\


\section{Simulation method and validation}\label{sec:Method}

\subsection{Numerical method}
Direct numerical simulations of single-phase flows are performed for an incompressible Newtonian fluid. A pseudospectral method is employed in the periodic directions (streamwise, $x$ and spanwise, $z$), and second-order finite differences are used for spatial discretization in wall-normal, $y$ direction. The solution is advanced in time by a third-order Runge-Kutta scheme. Incompressibility is achieved by correcting the pressure contribution, which is a solution of a Poisson equation. The fluid velocity and pressure fields are a solution of the continuity (\ref{eq:method_conti}) and momentum balance equations (\ref{eq:method_momen}) and (\ref{eq:method_drag_F}).

\begin{equation}\label{eq:method_conti}
\frac{\partial u_j}{\partial x_j}=0,
\end{equation} 

\begin{equation}\label{eq:method_momen}
\frac{\partial u_i}{\partial t} + u_j\frac{\partial u_i}{\partial x_j}=-\frac{1}{\rho_f} \frac{\partial p}{\partial x_i}+\nu \frac{\partial u_i}{\partial x_j \partial x_j} + \frac{1}{\rho_f} F_i,
\end{equation} 

\begin{equation}\label{eq:method_drag_F}
F_i^k = - \sum_{\mathclap{k_s=1}}^{8} ~ \sum_{\mathclap{n=1}}^{N_{k_s}}\frac{w_{k_s}^n}{\bigtriangleup V_{k_s}} f_{i}^n.
\end{equation} 

Using a projection technique, the body force in the momentum balance equation (\ref{eq:method_drag_F}) accounts for the particle momentum contribution to node $k$ of the fluid based on particles in all of the eight computational volumes ($k_s$) which share this node. $w_{k_s}^n$ is the linear geometric weight for each particle $n$ based on its distance from node $k$, and the inner summation is over all $N_{k_s}$ particles in the volume $k_s$.

Numerical simulations of particle trajectories and suspension flow dynamics are based on the standard Lagrangian point-particle approximation where the particle-to-fluid density ratio $r \equiv \rho_p/\rho_f \gg 1$ and the particle size is smaller than the smallest viscous dissipation scales of the turbulence. As a consequence of this and the low volume concentrations (a maximum volume fraction of $\overline{\Phi_{V}}$ less than $1 e-3$ is used in this study), only the Stokes drag force and two-way coupling have been incorporated \citep[see][]{balachandar2010turbulent}. The velocity of particle $n$ is governed in equation (\ref{eq:method_drag_f}) and particle trajectories are then obtained from numerical integration of the equation of motion (\ref{eq:method_motion}):

\begin{equation}\label{eq:method_drag_f}
\frac{d u^n_{p,i}}{dt}=\frac{f^n_i} {m_p}=\frac{1}{\tau_p}[1+0.15(\Rey^n _p)^{0.687}] (u^n_{f,i}-u^n_{p,i}),
\end{equation} 

\begin{equation}\label{eq:method_motion}
\frac{dx^n_i}{dt}=u^n_{p,i},
\end{equation} 

\noindent where $\tau _p = \rho _p {d_p} ^2 / 18\mu$ is the Stokes relaxation time of the particle, $m_{p}$ is the particle mass, and the particle Reynolds number $\Rey^n_p=\mid u^n_{f,i}-u^n_{p,i}\mid d^n_p / \nu $ is based on the magnitude of the particle slip velocity $(u^n_{f,i}-u^n_{p,i})$ and particle diameter $d_{p}^{n}$. In this work, the average $\Rey^n_p$ is between $0.125$ (r=80, two-way) and $3.25$ (r=8000, one-way), which is far smaller than the suggested maximum $\Rey_{p} \approx 800$ for the Stokes drag correction in equation (\ref{eq:method_drag_f}) \citep{schiller1933ber}. Thus the $\Rey_p$ correction to the Stokes drag is minimal in this study. In order to highlight the effect of particle response time, we do not consider gravitational settling. Other terms in the particle momentum equation \citep[see][]{maxey1983equation} are neglected since they remain small compared with drag when the density ratio $r \gg 1$. These neglected terms are also found to have less effect on the stability analysis in relatively low Reynolds number turbulence as demonstrated by \citet{klinkenberg2014linear}. Particle-particle collisions are not taken into consideration due to the low volume concentrations, and we exert a purely elastic collision between particles and the upper/lower walls. This purely elastic wall collision is commonly used in gas-solid turbulence \citep{li2001numerical,sardina2012wall,zhao2013interphasial}, however we have tested the restitution coefficient $|u^n_{p,init}/u^n_{p,final}|$ between $0.5$ and $1$ and do not observe significant changes to particle distributions or two-way coupling, consistent with \citet{li2001numerical}. Additional details and validation of the numerical scheme can be found in previous work for both one-way coupling and two-way coupling \citep{richter2013momentum,sweet2018gpu}.

\subsection{Single-phase flow in pCf}
\begin{figure}
\centering

\putfig{$(a)$}{\includegraphics[width=6.5cm]{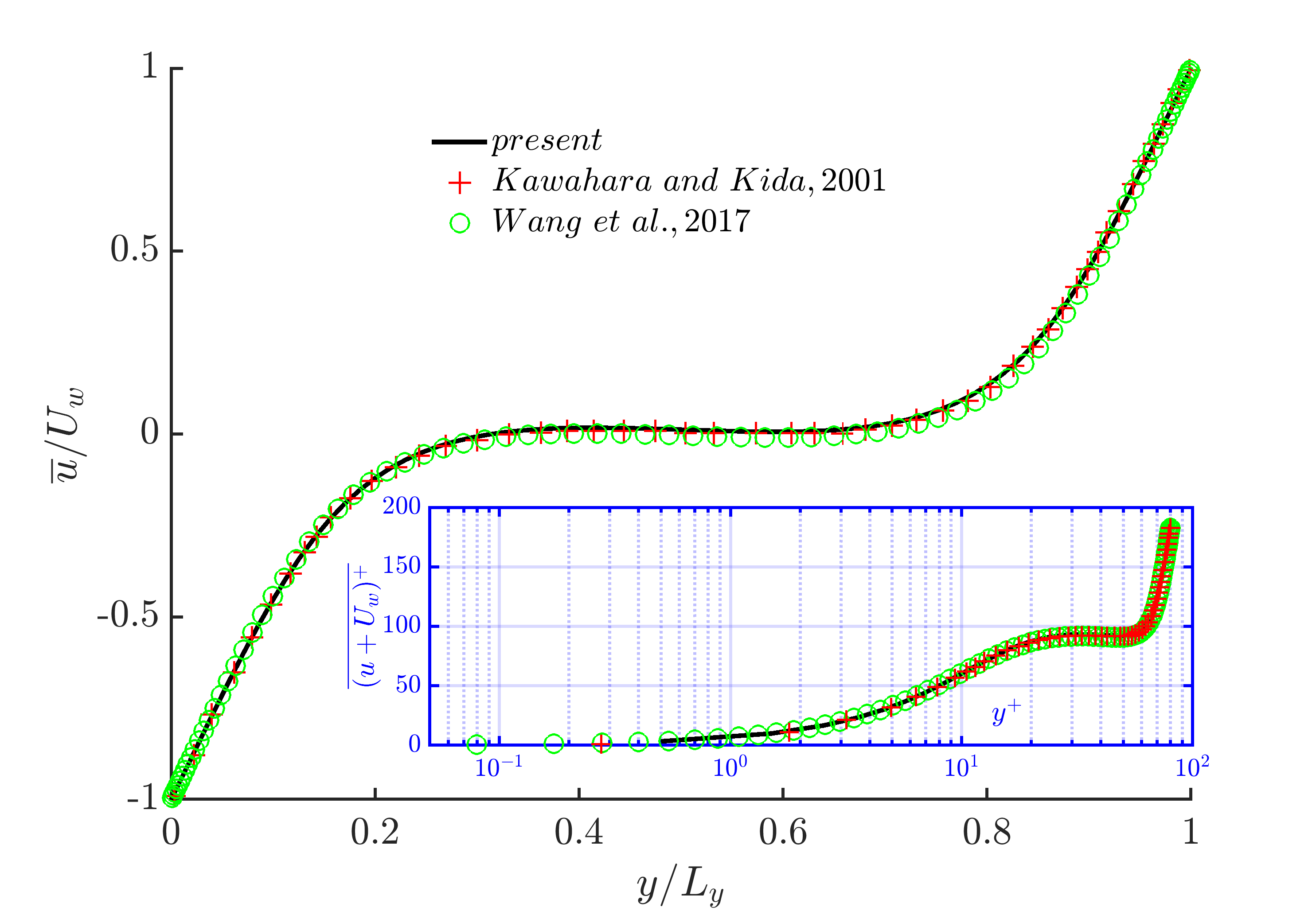}} \quad
\putfig{$(b)$}{\includegraphics[width=6.5cm]{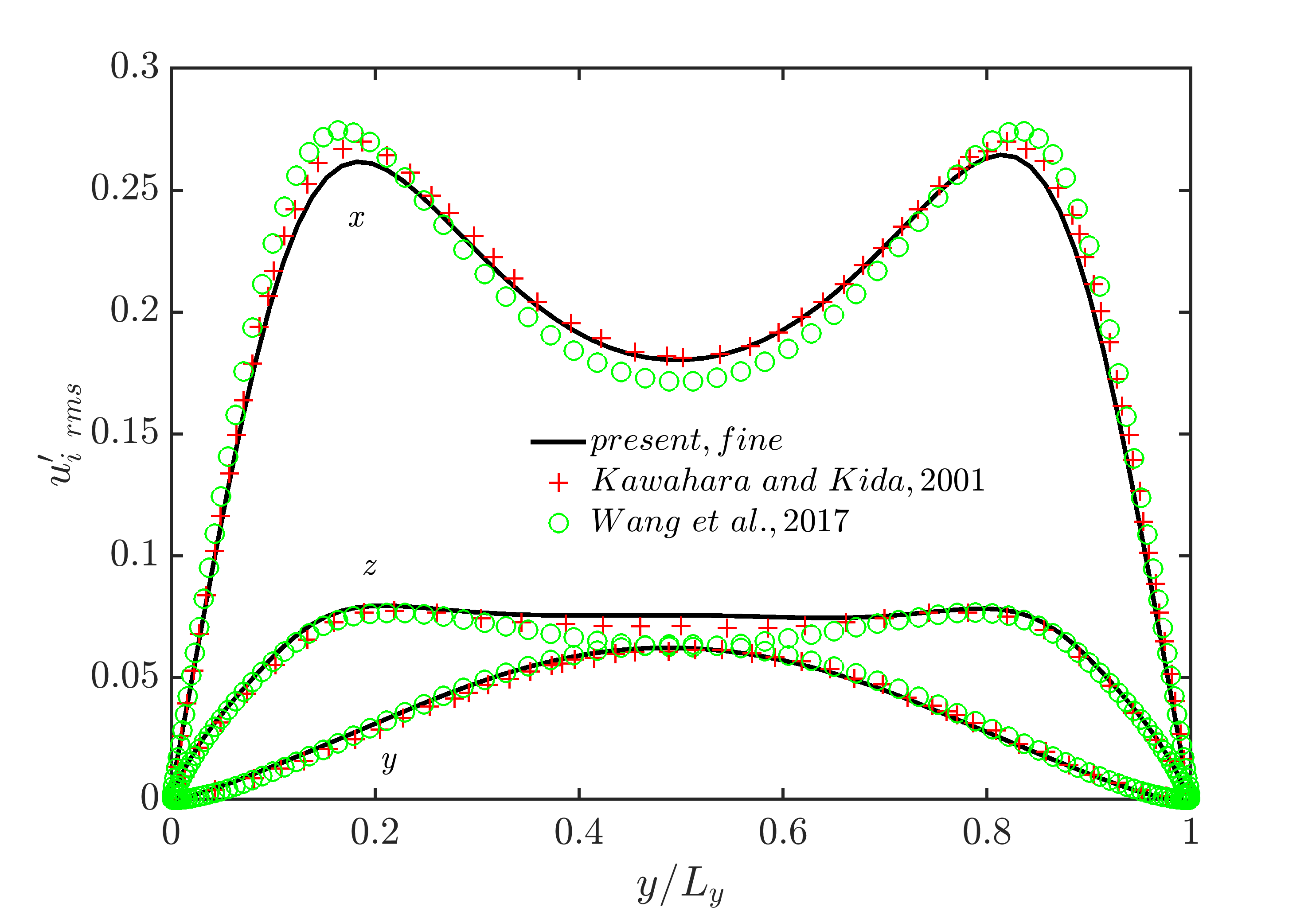}} \quad
  \caption{ Comparison between current study with published data from \citet{kawahara2001periodic} and \citet{wang2017modulation} in single-phase flow at $Re_b=400$: $(a)$ Mean velocity profiles; $(b)$ Fluctuation velocity RMS in three directions.}
\label{fig:Validation_Couette}
\end{figure}

Simulation of pCf turbulence in the miniunit configuration is used in the rest of this paper, at a relatively low Reynolds number (just above the onset of transition to turbulence). The grid used in this work is $N_x,~N_y,~N_z=32,~64,~32$ corresponding to a resolution in wall units of $\Delta x ^+ ,~ \Delta y ^+(wall,center) ,~ \Delta z ^+ \approx 6.8 ,~ (1,1.5) ,~ 4.7$. Figures \ref{fig:Validation_Couette}(a, b) show the mean velocity profile and velocity fluctuations in all three directions, in comparison with \citet{wang2017modulation} who used second-order centered finite volume method and \citet{kawahara2001periodic} who used a pseudospectral discretization in all three directions. We can see that the root-mean-square (RMS) velocity fluctuations from the present simulations agree well with \citet{kawahara2001periodic}, whereas there is a small discrepancy between the current work and \citet{kawahara2001periodic} in the streamwise and spanwise directions near the center region. This might be due to the slight difference of domain size in spanwise direction which confines the shape of two counter-rotating rolls in miniunit.
\subsection{Suspension flow configurations}\label{sec:Configuration}

\begin{table}
  \begin{center}
\def~{\hphantom{0}}
\begin{tabular}{ccccccccccc}
\multicolumn{11}{c}{\begin{tabular}[c]{@{}c@{}} Miniunit in plane Couette flow 
\\ 
\\ 
$L_y/d=80$; $L_x,~L_y,~L_z=2.75,~1.0,~1.88$\\
$N_x,~N_y,~N_z=32,~64,~32$ \\
$\Delta x ^+,~\Delta y ^+(wall,center),~\Delta z ^+= 6.8,~(1.0,1.5),~4.7$ \\
$~~~$
\end{tabular}} 
\\
$~~Case~~$   					& $r$   & $\overline{\Phi_m}$     & $u_{\uptau}$ & $\delta_\nu$ & $\Rey_ \uptau$         & $~L_y^+~$   & $~d^+~$     & $~\uptau_p^+~$     & $St_{turb}$  		   & $couple$
\\
$(\Rey_b-r-\overline{\Phi_v})$   	& $(\rho_p/\rho_f)$   &      & $(\times 10^{-2})$ & $(\times 10^{-2})$ &          &    &      &      &   		   & 
\\
\\
$500-0-0~~~$                    & --                	& --              		  & 4                & 1.25              & 40                      & 80        
& --            & --                 & --   				   & --
\\
Finite-size$~~~~$                 & 5                     & 1.0e-1             	   & 4.1             & 1.25              & 40                      & 80             & 4               & 4.44   			  & 0.056     			  & $FCM$
\\
$500-80-4~~$                    & 80                    & 3.2e-2             	   & 4.1             & 1.23              & 40.6                      & 81.2           & 1.01            & 4.58   			  & 0.056     			  & 2-way
\\
$500-80-10~$                    & 80                   & 8.0e-2             	   & 4.1             & 1.23              & 40.6                      & 81.2           & 1.01            & 4.58   			  & 0.056     			  & 2-way
\\
$500-500-4~$                    & 500                   & 2.0e-1             	   & 4.1             & 1.21              & 41.2                      & 82.5           & 1.03             & 29.5               & 0.347   			   & 2-way
\\
$500-900-4~$                    & 900                   & 3.6e-1             	   & 4.1             & 1.23              & 40.6                      & 81.2           & 1.02             & 51.6               & 0.625   			   & 2-way
\\
$380-900-4~$                    & 900                   & 3.6e-1             	   & 4.6             & 1.09              & 36.7                      & 91.7           & 0.92             & 41.9               & 0.499   			   & 2-way
\\
$500-5000-4$                   & 5000                   & 2.0             	       & 4.1             & 1.25              & 40                      & 80           & 1.01              & 278               & 3.47   			   & 1-way
\\
$500-8000-4$                   & 8000                   & 3.2             	       & 4.1             & 1.25              & 40                      & 80           & 1.01              & 444               & 5.56   			   & 1-way
\\
\end{tabular}
  \caption{Parameters of baseline numerical simulations. The Reynolds number $\Rey_b\equiv U_{w}h/\nu$ where $U_{w}$ is half of the relative velocity and $h=L_y/2$ is half of the gap between two walls. The friction Reynolds number is $\Rey_\tau\equiv u_{\uptau}h/\nu$ and the particle Stokes relaxation time is $\uptau _p \equiv \rho_p d^2/(18\rho_f \nu)$. The superscript ``+" is the dimensionless number based on viscous scale, where $\delta_\nu$, $u_{\uptau}$ and $\nu/ u^2_{\uptau}$ correspond to the viscous length scale, velocity scale, and time scale, respectively. FCM (Force Coupling Method) is a low order multipole expansion for capturing finite-size effect used by \citet{wang2017modulation}.}
  \label{tab:Table_1}
  \end{center}
\end{table}

Table \ref{tab:Table_1} contains several selected parameters for the statistical and modal analysis in this study (i.e., the cases used for all analyses but not including the transition tests of Section \ref{sec:transition test}). The case number indicates the bulk Reynolds number, density ratio, and particle volumetric concentration used in that particular simulation. Turbulence-laminar transition occurs near case $380-900-4$. \citet{hamilton1994streamwise} found a minimum threshold of streamwise vorticity circulation of the LSVs (below which the transition happens), and therefore we use this case to check if this threshold is modified by the presence of inertial particles. 

The Kolmogorov scale is difficult to determine and somewhat ambiguous at the low turbulence levels in these simulations, although we can estimate that the minimum dissipation scale is about $1.5$ (resp. $2.0$) times of viscous length scale at the wall (resp. in the center) \citep{pope2000turbulent}. The ratio of the particle to fluid time scale defines the Stokes number $St_{turb} \equiv \uptau _p / \uptau _f$, where $\uptau _f \equiv L_y/max(v'|w')$ is related to the turnover time scale of the large-scale vortices \citep{massot2007eulerian, wang2017modulation}. The estimation of $\uptau _f$ is about $25$ time units (time unit defined as $h/U_w$) from figures \ref{fig:RMS}(c,d). In order to gain insight into the finite-size effects on particle dispersion and turbulence modulation, we specifically select one group of parameters $~\uptau_p^+~$ and $St_{turb}$ to be nearly the same as the finite-size particles used in \citet{wang2017modulation}. There are two cases $500-5000-4$ and $500-8000-4$ with extremely high density ratios that are intended to help build a complete understanding of the effect of Stokes number on particle dispersion in the large-scale vortices in our model flow. These high density ratios are simulated with one-way coupling only since they fully laminarize the flow at $\Rey_b=500$.\

\section{Inertial particle response time effect on turbulent pCf}\label{sec:Statistics}

\subsection{Particle response time and mass loadings effect on transition}\label{sec:transition test}

\begin{figure}
\centering
\putfig{}{\includegraphics[width=15cm,trim={2.5cm 3cm 0 0}, clip]{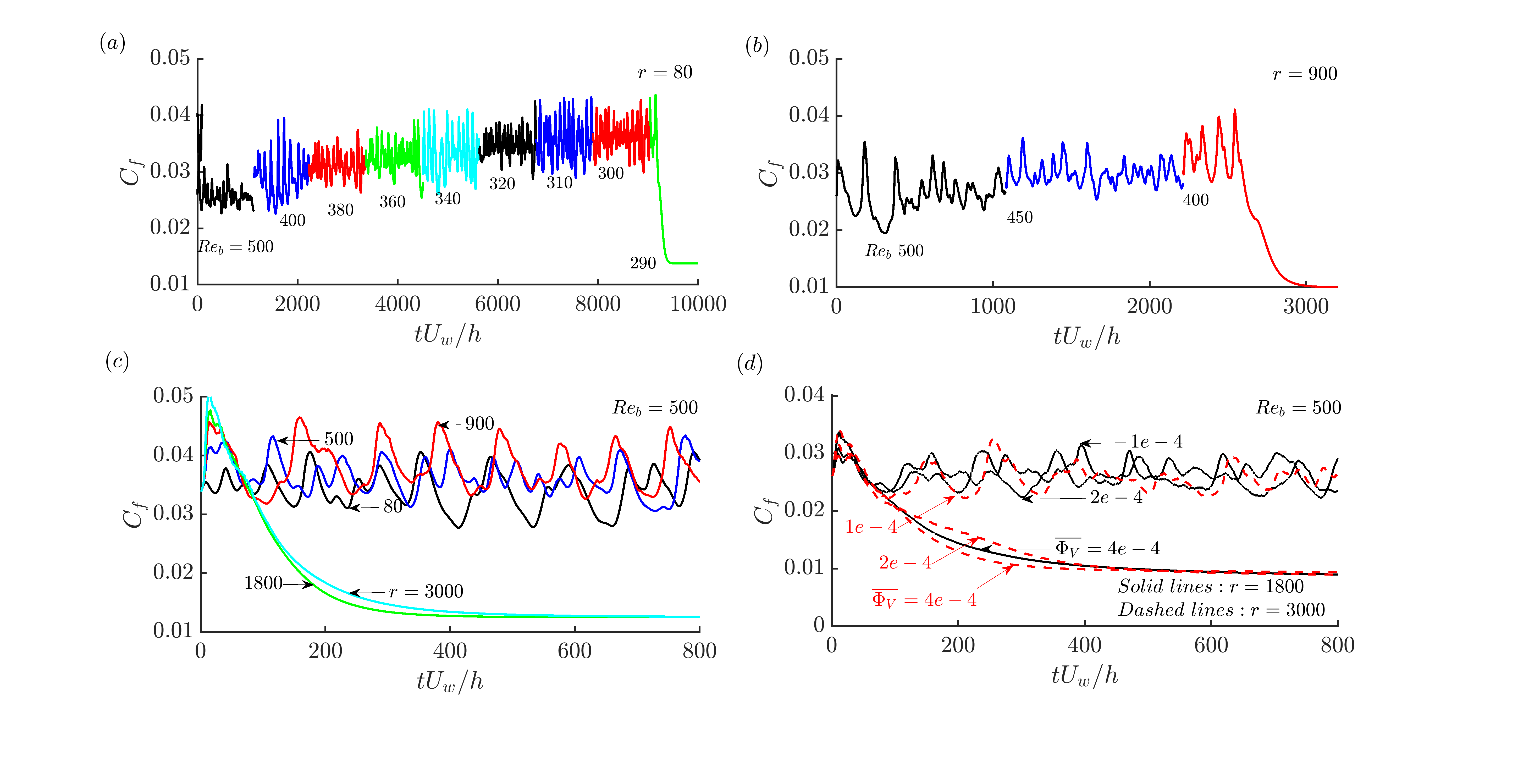}} \quad
  \caption{ Effect of inertial particles on the turbulence-to-laminar transition as indicated by the temporal evolution of $C_f$. This is demonstrated by gradually decreasing $Re_b$, density ratio ($r$) and bulk volumetric concentration ($\overline{\Phi_{V}}$) starting from a fully-turbulent simulation at $\Rey_b=500$. $(a)$ case $500-80-4$, decreasing $\Rey_b=500$ to $\Rey_b=290$; $(b)$ case $500-900-4$, decreasing $\Rey_b=500$ to $\Rey_b=400$; $(c)$ fixed $\Rey_b=500$ increasing density ratio $r=80$ to $r=3000$; $(d)$ fixed $\Rey_b=500$ and two density ratios $r=1800$ and $r=3000$, $\Phi_b$ decreasing from $4e-4$ to $1e-4$.}
\label{fig:Transition}
\end{figure}

While linear stability analysis is a key tool in understanding the effects of a dispersed phase on transition, it often must assume that the flow is uniformly (or with some other specified distribution) laden with particles, and that the flow is two-dimensional \citep{saffman1962stability,michael1964stability, dimas1998linear, rudyak1998instability}. Clearly, this type of analysis is fundamentally limited in uncovering nonlinear effects such as the regeneration cycle or inertial accumulation of particles \citep{squires1990particle,eaton1994preferential}, and theoretical predictions are difficult to develop. The interaction of inertial particles with coherent structures (e.g. LSSs and LSVs in pCf --- see figure \ref{fig:Dispersion}(a)) complicate matters and necessitate a fully nonlinear analysis. For this reason we examine the turbulent-to-laminar transition threshold from an empirical point of view, by simulating a fully-developed turbulent pCf flow experiencing $(i)$ successive reduction of the Reynolds number for several density ratios (figures \ref{fig:Transition}(a, b)), $(ii)$ successive increase of the particle density at a fixed Reynolds number (figure \ref{fig:Transition}(c)) and $(iii)$ successive increase of the particle mass loading for several density ratios at a fixed Reynolds number (figure \ref{fig:Transition}(d)). In each case, we are interested in the conditions under which the flow becomes laminar, though are objective is not to pinpoint precise values of the critical Reynolds numbers.

Transition of single-phase flow is observed at $\Rey_c\sim 320$ which is nearly same as in \citet{wang2018JFM} in the miniunit domain. The wall friction coefficient $C_f=2\overline{\uptau}_w/(\rho U_{w}^2)$ (summed on both walls) is used as an indicator of current flow regime. The same initial flow configurations were chosen from the single-phase flow simulations at $\Rey_b=500$ for all tests. The particles were then randomly seeded in the simulation domain. The two-phase flow simulations were integrated for at least $1200$ time units ($h/U_w$) before the Reynolds number was abruptly decreased, in order to accurately evaluate the transition threshold. The evolution in time of the wall friction coefficient is shown in figure \ref{fig:Transition} for the various tests. Figures \ref{fig:Transition}(a, b) show the successive reductions of the Reynolds number down to turbulent-to-laminar transition for $r=80$ ($\Rey_c \sim 290$) and $r=900$ ($\Rey_c \sim 400$) with same volumetric concentration of $\overline{\Phi_{V}} = 4 e-4$. We also perform the similar tests (not shown in this figure) for $r=200$ ($\Rey_c \sim 240$), $r=300$ ($\Rey_c \sim 260$), and $r=500$ ($\Rey_c \sim 310$). Figure \ref{fig:Transition}(c) shows the successive increase of the particle density up to the turbulence-to-laminar transition for $\Rey_b=500$ (transition occurs between $r = 900$ and $r=1800$) with same volumetric concentration of $\overline{\Phi_{V}} = 4 e-4$. Finally, figure \ref{fig:Transition}(d) shows the successive increase of the particle volume/mass loading for two density ratios $r=1800$ (transition occurs by $\overline{\Phi_{V}} = 4 e-4$) and $r=3000$ (transition occurs by $\overline{\Phi_{V}} = 2 e-4$).

Clearly, two limiting cases exist: at small density ratios (thus small Stokes number since $d_{p}$ is held constant) the flow experiences destabilization and the critical Reynolds number is lowered, and at high density ratios (high Stokes numbers) the particles tend to stabilize the turbulence and increase the critical Reynolds number. This is consistent with the linear stability prediction given by \citep{saffman1962stability}. For particles with high density ratios, the damping effect on turbulence varies monotonically with mass loading (again consistent with previous studies \citep{saffman1962stability, dimas1998linear, despirito2001linear}), at least in the dilute regimes considered here.  The turbulence attenuation effect is enhanced with an increased particle mass loading as seen in figure \ref{fig:Transition}(d). In pressure driven flow, \citet{klinkenberg2011modal} proposed a modified Reynolds number ($\Rey_m = (1+\overline{\Phi_m})\Rey_b$) for heavy particles via analysis of the standard Orr-Sommerfeld-Squire system. This effective Reynolds number can only predict turbulence damping (and an increase of the critical Reynolds number) and has a proportional increase with the particle mass loading. Comparing this estimate based on $Re_{m}$ to the present simulations, we obtain \\

$\Rey_c \sim 400-450$ for $r=900$ and $\overline{\Phi_V}=4 e -4$ (predicted $\Rey_m =435$); 

$\Rey_c \sim 500$ for $r=900$ to $1800$ and $\overline{\Phi_V}=4 e -4$ (predicted $\Rey_m =435-550$); 

$\Rey_c \sim 500$ for $r=1800$ and $\overline{\Phi_V}= 2 e -4$ to $4 e-4$ (predicted $\Rey_m =435-550$); 

$\Rey_c \sim 500$ for $r=3000$ and $\overline{\Phi_V}= 1 e-4$ to $2 e-4$ (predicted $\Rey_m =416-512$).\\

\noindent We therefore find that the transition Reynolds number found in the pCf laden with high-inertia particles ($St_{turb} > 0.5$, where turbulence is attenuated) is in the range of the estimates using the modified Reynolds number predicted by \citet{klinkenberg2011modal}.

\subsection{Conditional test of two-way coupling}

For pointwise particles, two-way coupling is realized by applying a corrected three-dimensional Stokes drag with Navier-Stokes equation as in equation (\ref{eq:method_drag_f}). The extra dissipation (the loss of kinetic energy due to fluid-particle interactions, see \citet{zhao2013interphasial}) plays the key role in providing a stabilizing effect. In order further investigate the key coupling components that lead to stabilization and turbulence attenuation, we perform a `conditional' test in this section.
   
\begin{figure}
  \centerline{\includegraphics[trim = 0mm 0mm 0mm 0mm,width=12.0cm]{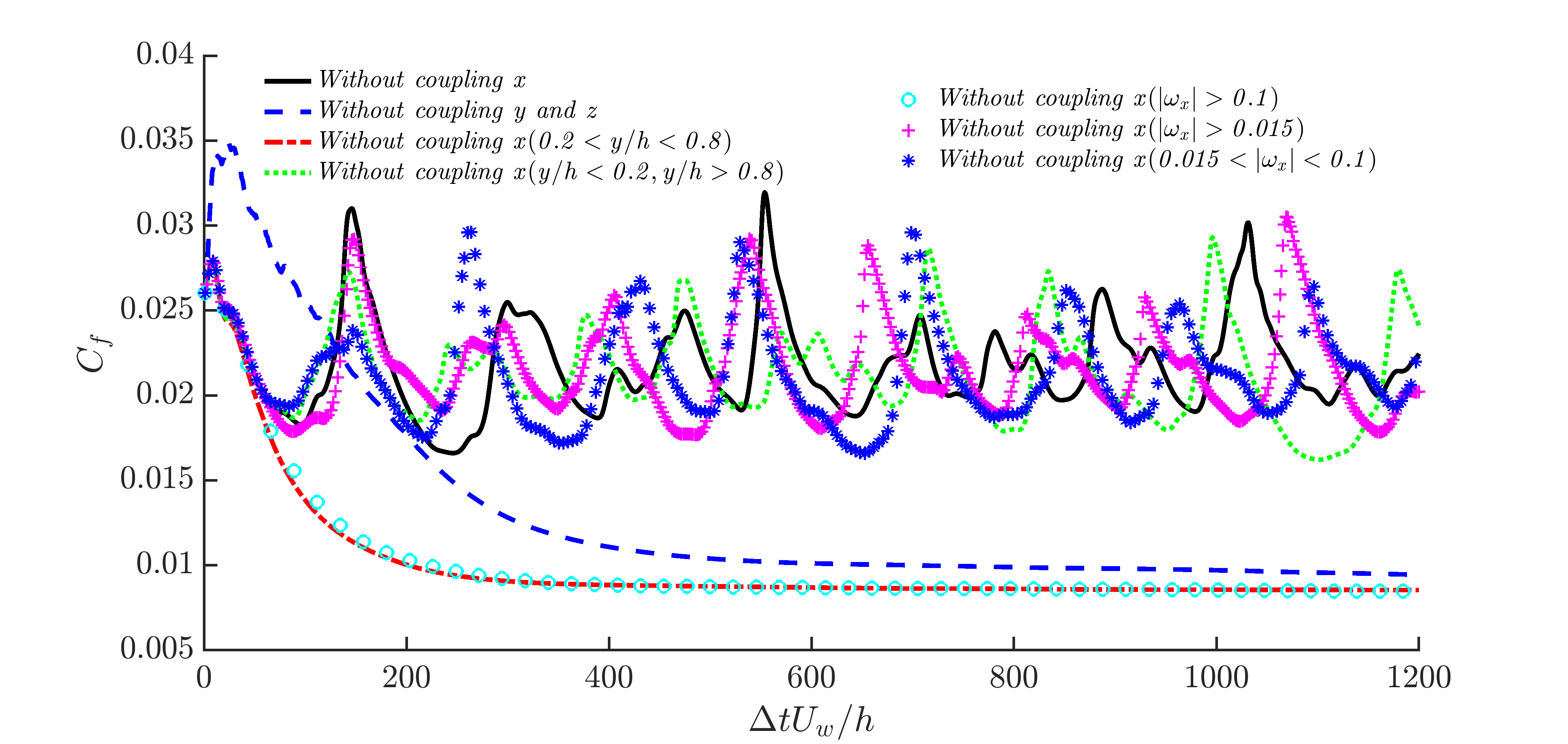}}
  \caption{Turbulence-laminar transition threshold indicated by the temporal evolution of $C_f$. A `conditional' test of sensitivity due to inertial particles drag force applied in different spatial regions in case $500-3000-4$.}
\label{fig:conditional_test}
\end{figure}

As shown in figure \ref{fig:conditional_test}, we begin with case $500-3000-4$ initialized from an unladen turbulent flow field at $\Rey_b = 500$ with randomly distributed particles. The flow decays into laminar flow with normal, full two-way coupling. By artificially removing the interphase coupling force in either the streamwise $x$ direction or both the $y$ and $z$ directions throughout the whole domain, we first find that streamwise coupling is the primary force that attenuates the turbulence (i.e. without $y$ and $z$ coupling the flow still transitions from turbulent to laminar). From here, the streamwise coupling is only removed (1) from the spatial region where the regeneration cycle is active ($0.2<y/h<0.8$) or (2) from the inactive region ($y/h<0.2$ and $y/h<0.8$). Turbulence is sustained by removing streamwise coupling in the region associated with the regeneration cycle, suggesting that it is streamwise coupling in this region which is responsible for laminarization. Further, we remove the streamwise coupling in different radial positions relative to the LSVs (strong streamwise vorticity $\mid \omega_x \mid>0.1$ is in center region of the LSVs, $\mid \omega_x \mid>0.015$ is a larger region containing $\mid \omega_x \mid>0.1$, and $0.015<\mid \omega_x \mid<0.1$ is an annulus formed by subtracting the area $\mid \omega_x \mid>0.1$ from $\mid \omega_x \mid>0.015$). This analysis, shown in figure \ref{fig:conditional_test} as blue stars, demonstrates that streamwise coupling, specifically in the region associated with the annulus given by $0.015<\mid \omega_x \mid<0.1$ and in the region associated with the regeneration cycle, should be considered as the most sensitive region which might be altered by the inertial particles, leading to turbulence attenuation.

\begin{figure}
\centering
\putfig{}{\includegraphics[width=15cm,trim={4cm 1.5cm 0 2cm}, clip]{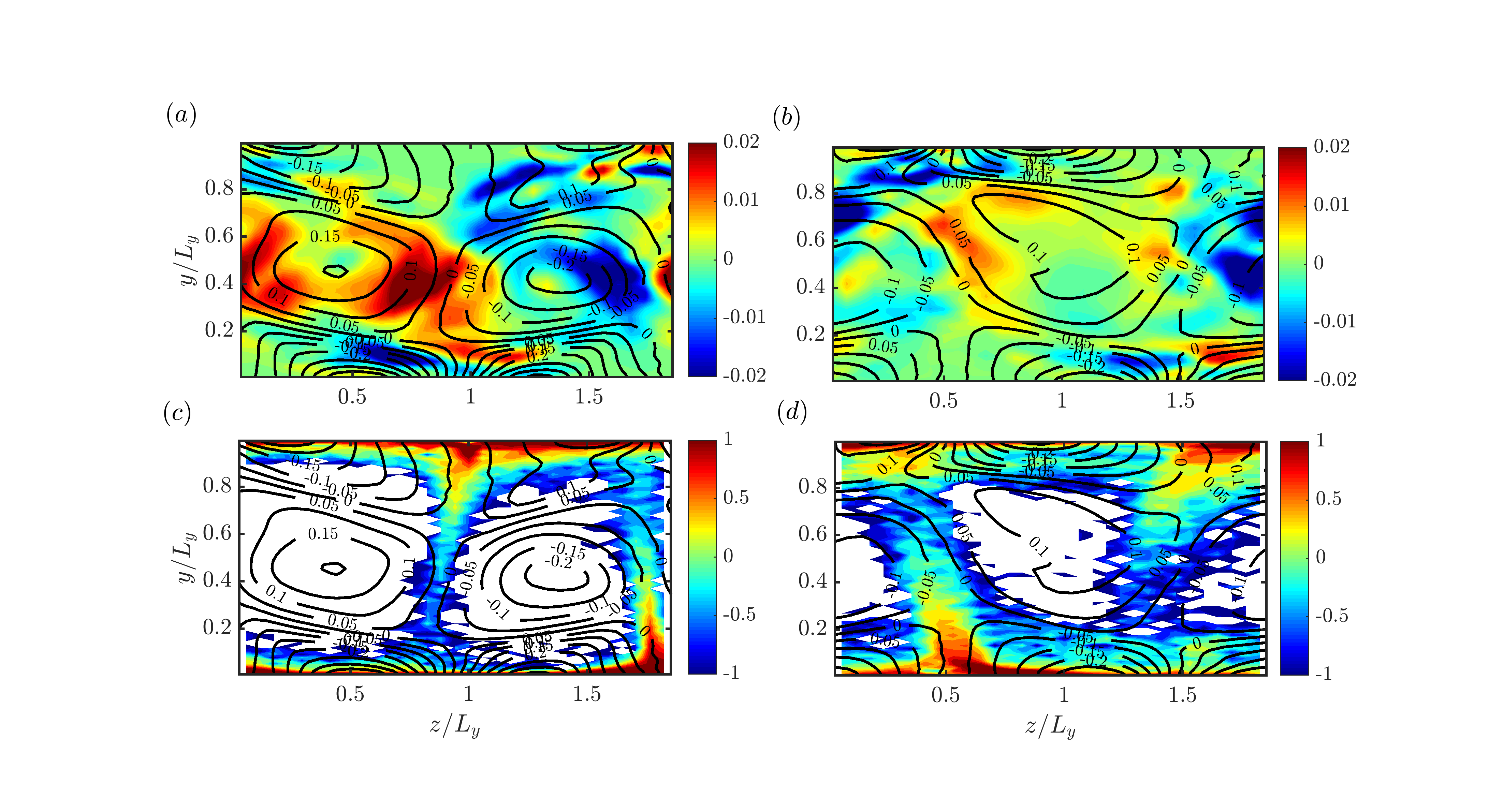}} \quad
  \caption{$Top~panel$: contour of temporal average of streamwise vortex stretching term in one single cross-section (y,z plane) within $50$ time units and isolines of temporal and streamwise average of streamwise vorticity within $50$ time units. $Bottom~panel$: contour (logarithmic scale) of temporal and streamwise average of normalized concentration by the bulk value and isolines of temporal and streamwise average of streamwise vorticity within $50$ time units. $(a,c)$: case $500-500-4$; $(b,d)$: case $500-900-4$.}
\label{fig:conditional_test_contour}
\end{figure}

To visualize this stabilization process that occurs in the annulus between $0.015<\mid \omega_x \mid<0.1$, the top panel in figure \ref{fig:conditional_test_contour} shows the contour of streamwise vorticity stretching $\omega_x \partial u/ \partial x$. This is a streamwise-dependent quantity, so we time average (50 time units) at a single cross-section at $x=L_x/2$. The bottom panel in figure \ref{fig:conditional_test_contour} shows the contour of the normalized particle concentration with respect to the bulk concentration at the same location and averaged over the same time. The isolines in all panels are the streamwise averaged streamwise vorticity ($\overline{\omega_x}(y,z)$). The left panel and right panel show cases $500-500-4$ and $500-900-4$, respectively. It is clear that strong vorticity stretching happens in the range of $0.015<\mid \omega_x \mid<0.1$, which is a key component of the self-sustaining regeneration cycle. The intensity of this stretching is reduced in case $500-900-4$ (contour in figure \ref{fig:conditional_test_contour}(b)) compared to that of case $500-500-4$ (contour in figure \ref{fig:conditional_test_contour}(a)). Simultaneously, we find that there are more particles present in the range of $0.015<\mid \omega_x \mid<0.1$ in case $500-900-4$ (figure \ref{fig:conditional_test_contour}(d)) than in $500-500-4$ (figure \ref{fig:conditional_test_contour}(c)).

As a result, we associate the turbulence attenuation to the streamwise coupling of high-inertia particles (e.g. $St_{turb}=0.625$), and their presence in the range of $0.015<\mid \omega_x \mid<0.1$ is of key significance since this is the region which has the strongest streamwise vortex stretching effect. Below, we argue that the streamwise vortex stretching and the large-scale vortices are the key sub-steps in the regeneration cycle, and coupled to one other. The preferential presence of particles in this streamwise vortex stretching region is an important phenomenon which alters the regeneration cycle and thus the transition from turbulent to laminar flow, and has been obseved at higher Reynolds numbers \citep{lee2015modification}.\\

\subsection{Particle distribution and velocity profile}

For inertial pointwise particles, their spatial distribution is determined through the drag force exerted from large-scale turbulent structures to the particles. Due to inertia, heavy particles tend to be expelled from the vortex center to the low-vorticity but high strain-rate regions \citep{druzhinin1998direct}. As argued above, this process plays a key role in turbulence modulation since particles tend to accumulate in regions associated with streamwise vortex stretching. In a similar study focused on neutrally buoyant finite-sized particles, \citet{wang2018JFM} found that particles accumulating in the streaks tend to enhance the turbulence whereas particles accumulating in the large-scale rolls hardly modify turbulence levels.

\begin{figure}
\centering
\putfig{$(a)$}{\includegraphics[width=13.5cm]{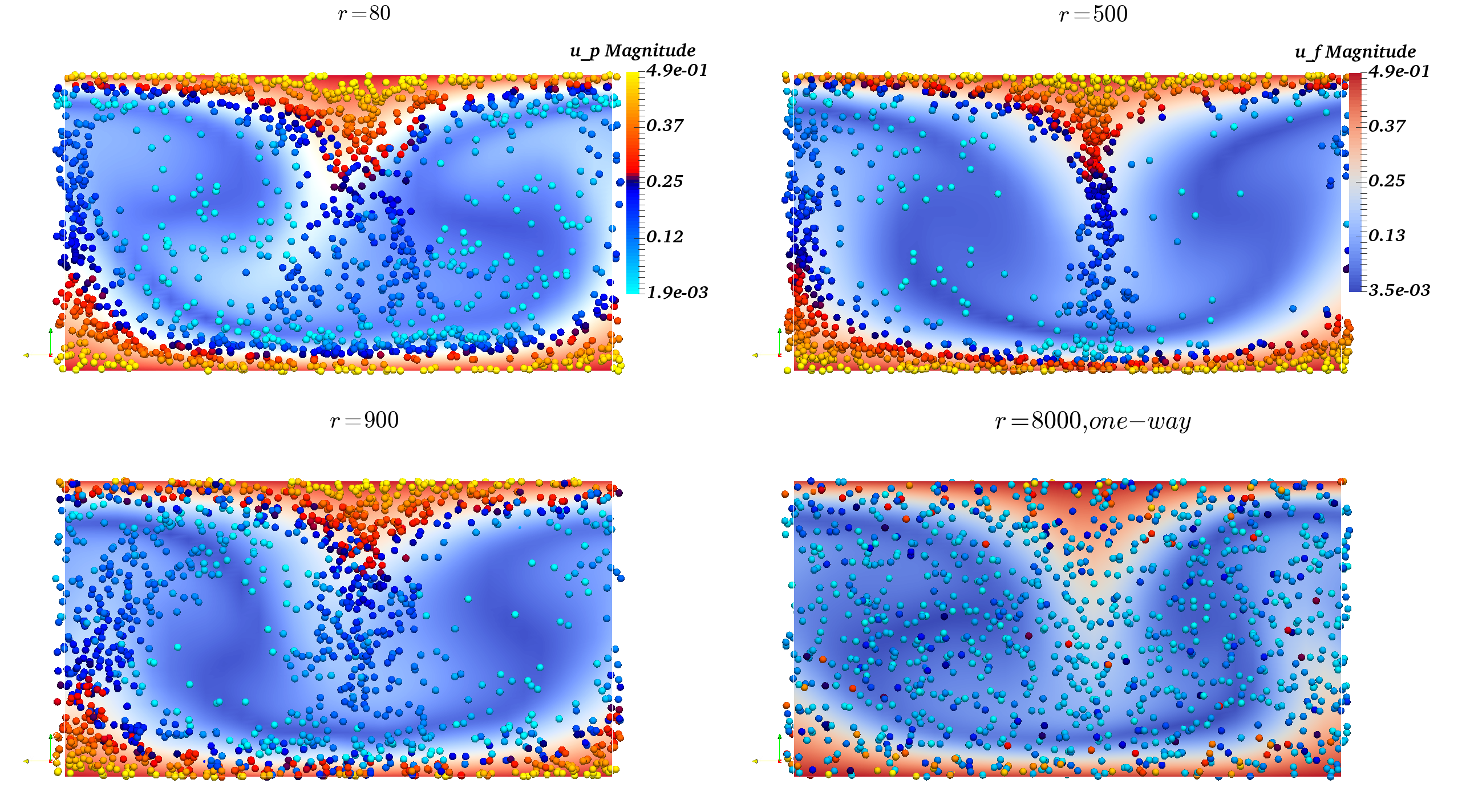}} \quad
\putfig{$(b)$}{\includegraphics[width=6.5 cm]{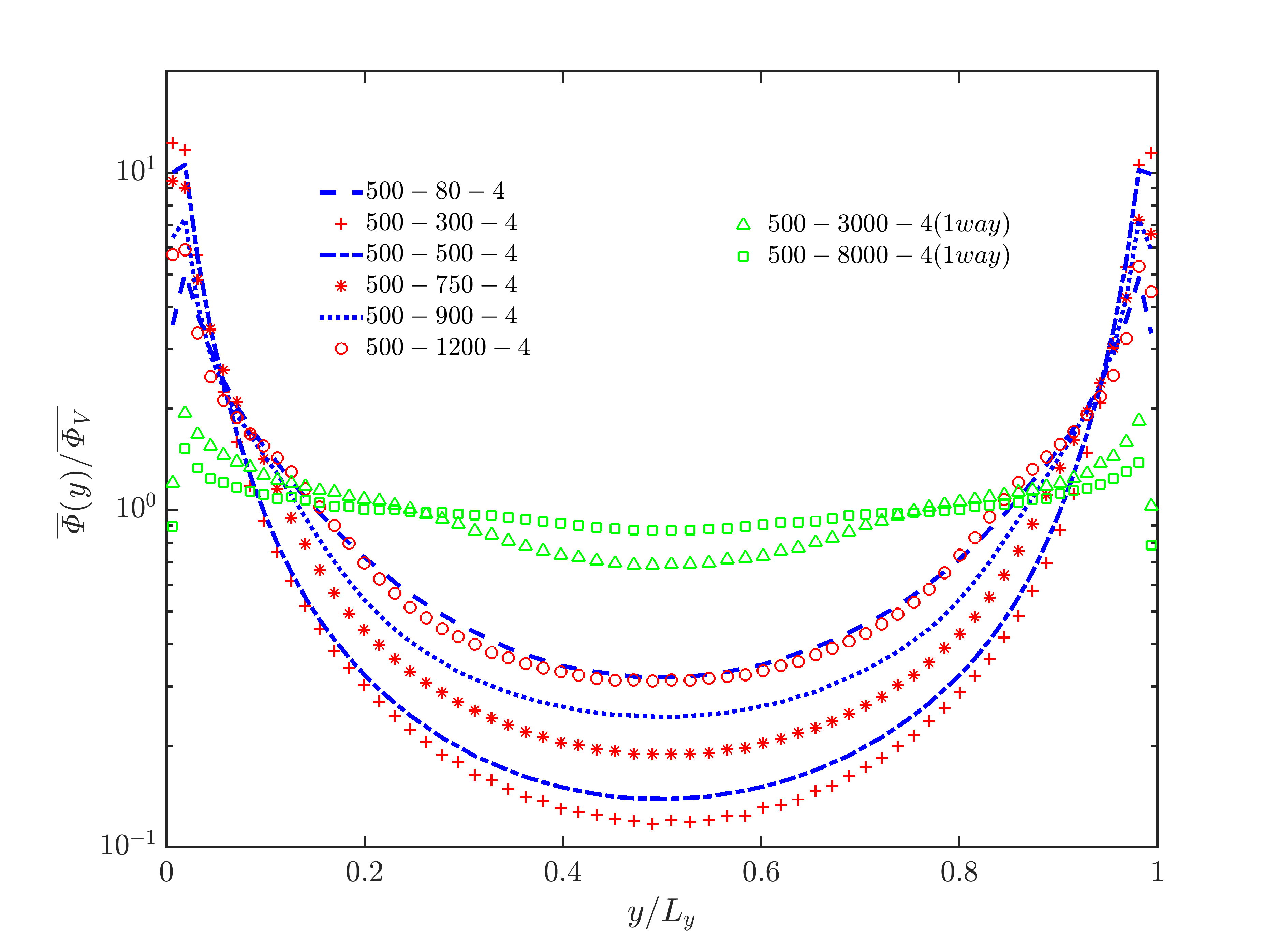}} \quad
\putfig{$(c)$}{\includegraphics[width=6.5 cm]{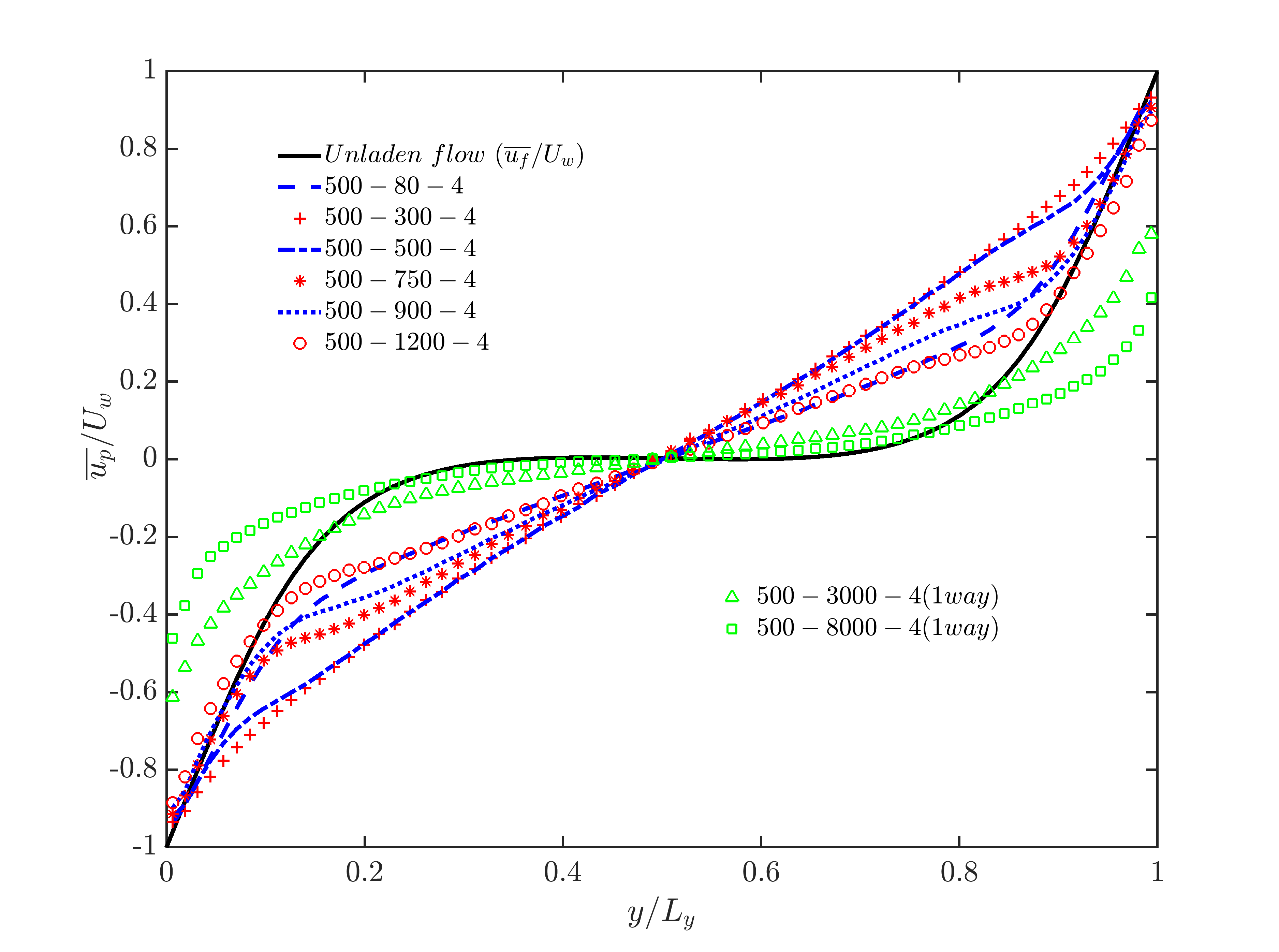}} \quad
  \caption{ $(a)$ Contours of the magnitude of the streamwise flow velocity (colorbar in upper left panel) and instantaneous particle positions projected onto the $(y,z)$ cross-section when the LSS is strongest before breakdown, colored according to magnitude of particle streamwise velocity (colorbar in upper right panel). The particle size shown in the figures is magnified four times for better visualization. $Top-left$: case $500-80-4$; $Top-right$: case $500-500-4$; $Bottom-left$: case $500-900-4$; $Bottom-right$: case $500-8000-4$ (one-way coupling). $(b)$ Mean volume concentration normalized by the bulk concentration; $(c)$ Mean streamwise velocity scaled by $U_w$.}
\label{fig:Dispersion}
\end{figure}

Figure \ref{fig:Dispersion}(a) shows an instantaneous particle distribution over a $(y,z)$ cross-sectional plane with different density ratios ($r=80-8000$) at the same Reynolds number ($\Rey_b=500$) and volumetric concentration ($\overline{\Phi_v}=4 e-4$) at a point in time corresponding to the strongest LSSs during the regeneration cycle. From left to right, top to bottom, it is clear that particles with low to moderate inertia tend to accumulate in the LSSs whereas particles with a higher inertia distribute more homogeneously and spread throughout the Couette gap. Along the same lines, particles with density ratio from $r=300$ to $500$ are mostly trapped inside the LSSs, even during the streak breakdown process (not shown). Either lower (e.g. $r=80$) or higher (e.g. $r=900$) than this density ratio, the particles can stray from the LSSs regions, consistent with known behavior of inertial particles \citep{maxey1987motion, druzhinin1998direct}. In a Taylor-Green Vortex (TGV) setup, \citet{massot2007eulerian} proposed a threshold of the ratio between particle response time with the TGV turnover time scale. Below the threshold particles will stay inside this cell whereas particles tend to move to the other cells above the value. In Appendix A, we numerically obtained the single particle trajectory in TGV flow with the same particle time scale corresponding to particles in the present simulated turbulent LSVs. The case $500-80-4$ has a low $St_{turb} \approx 0.056$ causing the particles to behave more as tracers and stay in one LSV, whereas case $500-8000-4(1way)$ with a high $St_{turb} \approx 5.56$ leads to particles which cannot follow the streamlines and cross the LSVs after being ejected by ejection events. Particles with intermediate Stokes numbers, as in case $500-500-4$ with $St_{turb}\approx 0.35$, are expelled from the LSVs and become trapped inside the LSSs.

This non-monotonic change in particle distribution can be seen in figure \ref{fig:Dispersion}(b), where the normalized particle volume concentration is shown. The concentration first decreases (resp. increases) with increasing density ratio from $500-80-4$ to $500-300-4$ in the center region (resp. near-wall region), while it has an opposite tendency when further increasing the density ratio from $500-300-4$ to $500-1200-4$. At higher density ratios with two-way coupling, transition to laminar flow occurs in the present simulations. At higher flow Reynolds number ($\Rey_b=2025$), the increase of particle concentrations in the center with increasing particle response time ($\uptau_p^+ \approx 90$ corresponding to $St_{turb} \approx 1.06$) is also observed by \citet{richter2013momentum}. For the sake of highlighting the long particle response time effect on the particle distribution, we also show cases $500-3000-4$ and $500-8000-4$ with one-way coupling, where we can find a nearly homogeneous concentration profile because particles cannot follow the streamlines.

Despite the slip velocity between the particulate phase and fluid phase, the shape of the mean particle velocity profile is mainly determined by the carrier phase. In particular, the mean particle streamwise velocity will reflect the local mean fluid velocity profile of the structure it is contained within (e.g. LSSs or LSVs). \citet{hamilton1995regeneration} has shown that for the miniunit configuration, the characteristic `S' shape of the mean velocity profile in pCf is governed primarily by the LSVs in single-phase flow. Figure \ref{fig:Dispersion}(c) compares the mean fluid velocity (solid black line) to the mean particle velocities $\overline{u_p}/U_w$ of particles with varying density. There is a clear distinction between the nearly linear (resp. `S') shape of the mean particle velocity curve for $500-300-4$ (resp. $500-3000-4$(1way)) if the particles are trapped in the LSSs (resp. across the whole domain). Again, this effect is non-monotonic with particle inertia. For cases $500-3000-4$(1way) and $500-8000-4$(1way), the qualitative shape of the mean particle velocity is similar to mean fluid velocity in single-phase flow; this discrepancy is due to the slip induced by the unmatched time scale between particles response time with the LSV turnover time.

\subsection{Turbulence intensity}

\begin{figure}
\centering
\putfig{}{\includegraphics[width=14.5cm,trim={4cm 0cm 0 0}, clip]{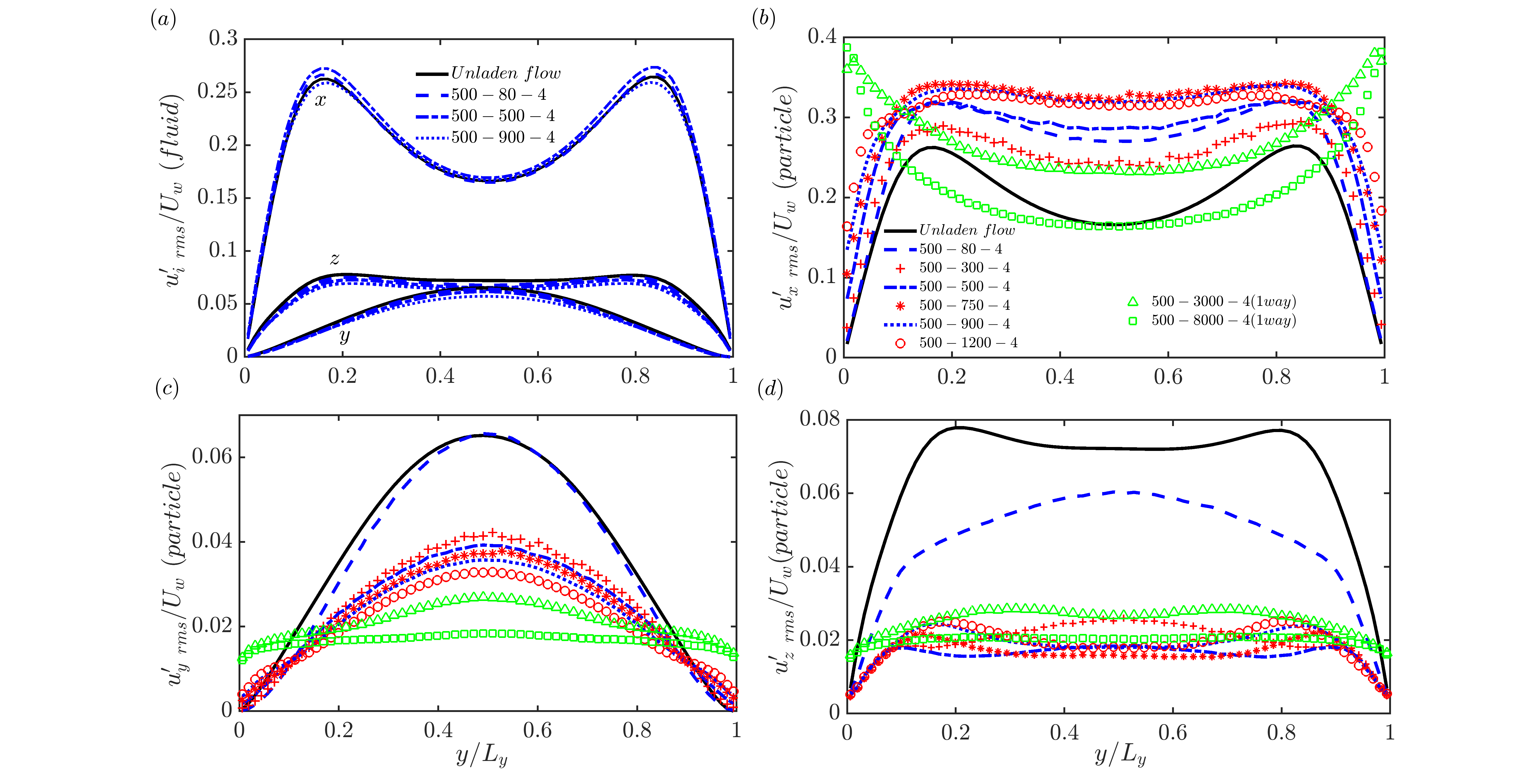}} \quad
  \caption{ RMS fluctuation velocity in cases with density ratios. $(a)$ Fluid phase in three directions; $(b-d)$ Particulate phase in three directions: $(b)$ Streamwise direction fluctuation; $(c)$ Wall-normal direction fluctuation; $(d)$ Spanwise direction fluctuation. In $(b-d)$, fluid phase in single-phase is plotted as a reference.}
\label{fig:RMS}
\end{figure}

The RMS velocity fluctuations for the various cases are plotted in figure \ref{fig:RMS}. Figure \ref{fig:RMS}(a) shows the RMS velocities in all three directions for the fluid phase, where it is apparent that $u^\prime_{f~rms}$ is nearly unchanged whereas in the core region the $v^\prime_{f~rms}$ and $w^\prime_{f~rms}$ decrease slightly with increased particle inertia. The particulate RMS velocity fluctuations are shown in figures \ref{fig:RMS}(b-d); the single-phase velocity fluctuations are also plotted as a reference (solid black lines). As noted by \citet{yu2016finite}, particles moving away from the wall are associated mainly with ejections while particles moving towards the wall are associated with sweeps. As seen above, inertial particles with low to moderate Stokes numbers tend to remain in the high strain rate region (see figure \ref{fig:Dispersion}) --- regions associated with high $u^\prime_{f~rms}$. This results in high values of $u^\prime_{p~rms}$ across the whole Couette gap which can be seen in figure \ref{fig:RMS}(b). Thus the increase of particulate streamwise turbulent kinetic energy compared with single-phase flow is due to the accumulation of particles in the LSS which contains high streamwise turbulent kinetic energy.

The effect is opposite for $v^\prime_{p~rms}$ and $w^\prime_{p~rms}$. The fluid velocity fluctuations $v^\prime_{f~rms}$ and $w^\prime_{f~rms}$ are high in the outer regions of the LSVs (not shown here). Therefore when the inertial particles collect in high strain rate regions (intermediate Stokes numbers), their spanwise and wall-normal velocity fluctuations are much smaller than the fluid average. At low Stokes numbers, long residence times in the LSVs allow particles to gain wall-normal and spanwise kinetic energy (figures \ref{fig:RMS}(c) and \ref{fig:RMS}(d)). At high Stokes number, particles tend to again distribute homogeneously throughout the whole domain. Particles move across LSVs, but their inability to quickly adjust their velocity results in suppressed values of $v^\prime_{p~rms}$ and $w^\prime_{p~rms}$. 


As a final note, we have compared the particle distribution, mean velocity and RMS velocity fluctuations between case $500-80-4$ and a simulation with finite-size particles from \citet{wang2017modulation}; both have the same particle response time. We find that even for the same Stokes number, finite-size particles collect more in the central region and that their mean velocity and fluctuations are similar to the fluid phase. This difference is due directly to finite-size effects which should be considered for physical systems where particle diameters are larger than the dissipation scales of the flow.



\section{Modal analysis of the regeneration cycle modulation} \label{sec:regeneration_cycle} 

Turbulence regeneration mechanisms in wall turbulence involve three-dimensional, multiscale structures, and the relevant nonlinear dynamics involved are neglected when performing linear stability analyses on laminar-to-turbulent transition and streak stability. Therefore to leverage the present nonlinear simulations, modal analysis is used to determine the natural mode shapes and frequencies in this miniunit pCf system --- this strategy significantly simplifies the range of temporal and length scales in the turbulent flow, thereby highlighting the essential features (e.g. LSSs and LSVs) in the regeneration cycle. To start, it is helpful to provide a list of relevant quantities which are useful in defining the various stages/structures of the regeneration cycle, which will be further analyzed in this section. \\ 

\noindent 1. The turbulent kinetic energy contained in the streaks and its corresponding Fourier modes yield information about the contributions from various scales in the flow. We define $\mathcal{M}(k_{x} = m\alpha,k_{z} = n\beta)$ as the vertically integrated modal RMS velocity modes in the two periodic directions ($x$ and $z$), following \citet{hamilton1995regeneration}:

\begin{equation}\label{eq:fft_LSS} 
\mathcal{M}(m\alpha ,n\beta) \equiv  \left\lbrace \int ^{L_y}_{0} \left[ \widehat{u^\prime}^2(m\alpha,y,n\beta) + \widehat{v^\prime}^2(m\alpha,y,n\beta)+\widehat{w^\prime}^2(m\alpha,y,n\beta) \right] dy \right\rbrace ^{1/2},
\end{equation}

\noindent where $(\alpha,\beta)$ are the fundamental wavenumbers in the streamwise and spanwise directions (defined as $(2\pi/L_x,2\pi/L_z)$), and $m$ and $n$ are integers. In principle, each mode $(m\alpha,n\beta)$ provides information about specific structures in the flow. For instance in the miniunit domain (note that this may not necessarily be true in full domains), \\

$\mathcal{M}(0,\beta)$ represents the LSSs since it is the zeroth mode in the streamwise direction, and any mode $(0,n\beta)$ with $n\neq0$ is an $x$-independent structure;

$\mathcal{M}(\alpha,0)$ represents meandering streaks where the other $\mathcal{M}(m\alpha, n\beta)$ with $n\neq0$ modes are very weak.\\

\noindent 2. Circulation， represents the intensity of the LSVs in the miniunit. \citet{hamilton1994streamwise} and \citet{hamilton1995regeneration} stipulated that over one cycle, vortices must have a maximum circulation above a given threshold in order to produce an LSS through the lift-up process. Thus this integrated quantity provides a good measure of whether or not this essential process can occur. The circulation of the streamwise vorticies of mode zero in the streamwise direction ($x$-independent), and mode $n\neq0$ in the $z$ direction (z-dependent) is given by

\begin{equation}\label{eq:fft_LSV}
\centering
\mathcal{C}(0,n\beta)  \equiv \int ^{L_y}_{0}~ \hat{\omega}_x(0,y,n\beta) d \bf{S(n)},
\end{equation}

\noindent where $\mathbf{S(n)} \equiv dy \cdot L_z/n$ is the transverse surface with $y$ varying from $0$ to $L_y$ and $z$ varying from $0$ to $2\pi/(n\beta)$ for $n \neq 0 $. Based on our calculations, the maximum circulation in the miniunit domain always corresponds to $n=1$, and therefore we define circulation as $\mathcal{C}(0,\beta)$ and use this measure to stand for the intensity of LSVs.\\

\noindent 3. Lift-up: in the buffer layer, elongated streaks form on both sides of an LSV (see figure \ref{fig:Dispersion}(a)). The so-called lift-up effect has been identified as a very robust mechanism for the generation of streaky motions both in transitional and turbulent flows \citep{ellingsen1975stability, hamilton1995regeneration, brandt2014lift}. Fluid in the near-wall region is lifted away from by the longitudinal vortical structures into a region of higher-speed fluid (so-called ejection), producing a low-speed streak. Simultaneously on the other side of the vortex, high-speed fluid is pushed towards the wall (so-called sweep), creating a high-speed streak. Consequently in shear flows, the main linear mechanism for transient disturbance growth is the lift-up effect that produces low- and high- speed streaks in the streamwise velocity \citep{ellingsen1975stability}. \citet{bech1995investigation} stated that the inner shear layer is formed via the lift-up of low-speed streaks from the viscous sublayer. Once this occurs, the shear layers are coupled to an instantaneous velocity profile with an inflectional character, and they have been observed to become unstable and break up into chaotic motion, so called `bursting'. Specifically in Fourier space, \citet{hamilton1995regeneration} has shown the term most responsible for extracting energy from the mean shear flow is given by
\begin{equation}\label{eq:fft_Liftup}
\centering
\mathcal{L}(0,n\beta)  \equiv \int ^{L_y}_{0}~\widehat{v^\prime}(0,y,n\beta) \frac{\partial U(y)}{\partial y}~dy,
\end{equation}
where $\widehat{v^{\prime}}$ is the wall-normal fluctuation velocity in Fourier space and ${\partial U(y)}/{\partial y}$ is the gradient of the mean steamwise velocity.\\



\noindent 4. Steamwise vortex stretching: during streak breakdown, nonlinear interactions reinforce the streamwise LSVs, leading to the formation of a new set of streaks, and completing the regeneration cycle. \citet{hamilton1995regeneration} proposed that the strengthening of the vortices is due to interactions among the $\alpha$-modes, which grow during the streak breakdown. \citet{schoppa2002coherent} suggested that the vortex formation is inherently three-dimensional, with direct stretching (inherent to streak $(x,z)$-waviness) of near-wall $\omega_x$ sheets leading to streamwise vortex collapse. They provided insights into the dynamics of near-wall vortex formation through the inviscid equation for streamwise vorticity: 

\begin{equation}\label{eq:vorticity_dynamic}
\frac{\partial{\omega_x}}{\partial{t}}=\underbrace{-u\frac{\partial{\omega_x}}{\partial x}-v\frac{\partial{\omega_x}}{\partial y}-w\frac{\partial{\omega_x}}{\partial z}}_\textit{advection} +\underbrace{{\omega_x}\frac{\partial u}{\partial x}}_\textit{stretching}+\underbrace{\frac{\partial v}{\partial x}\frac{\partial u}{\partial z}-\frac{\partial w}{\partial x}\frac{\partial u}{\partial y}}_\textit{tilting} ,
\end{equation}
In fully developed turbulence, the greatest contribution, in magnitude, to the temporal evolution of the vorticity $\partial \omega_x / \partial t$ is related to the tilting term \citep{sendstad1992near}.  However, \citet{schoppa2002coherent} have shown that this term mainly contributes to the thin tail of the near-wall $\omega_x$ layer, and is not responsible for $x$-independent streamwise vortex formation ($(0,\beta)$ mode in the miniunit). Instead, vortex formation is dominated by stretching of streamwise vorticity. The local $\omega_x$ stretching is sustained in time and is mainly responsible for the vortex collapse, whose location coincides with the $+\omega_x \partial u / \partial x$ peak. The meandering of streaks provides the generation of $\partial u / \partial x$, and then direct stretching of positive and negative $\omega_x$ occurs in regions where $\partial u / \partial x$ is generated across the wavy streak flanks during the streak breakdown process. The stretching term is active only during the peaks of the cycle when local three-dimensionality is induced after streak breakdown \citep[see][]{jimenez1991minimal}. In Fourier space, the stretching term can be written as 

\begin{equation}\label{eq:fft_stretching} 
\mathcal{S}(m\alpha,n\beta)  \equiv \int ^{L_y}_{0}~ \widehat{\omega_x} (p\alpha,y,q\beta) \frac{\partial \widehat{u} (r\alpha,y,s\beta)}{\partial x} ~dy,
\end{equation}

\noindent where the total time rate of change of streamwise vorticity in the $(m\alpha,y,n\beta)$ mode is the summation of terms over all values of $p,~q,~r,~s$ such that $p+r=m$ and $q+s=n$. The contribution to the LSV with $(0,\beta)$ comes from wavenumber combinations which satisfy $p+r=0$ and $q+s=1$. For the streamwise direction, \citet{hamilton1995regeneration} demonstrated that $p= \pm 1$ and $r= \mp 1$  are the dominant terms producing additional streamwise vorticity in the right places where the higher $x$-wavenumber modes are negligible. It is difficult, however, to specify a single pair of modes combining both $x$-wavenumber and $z$-wavenumber which are dominant. 

Therefore, we consider all pairs $p= \pm 1$ and $r= \mp 1$ satisfying the condition $m=p+r=0$ in streamwise direction and $q=-2$ to $3$ and $s=3$ to $-2$ satisfying the condition $n=q+s=1$ in spanwise direction (a total $12$ pairs of $x$ and $z$ wavenumbers). A discrete Fourier transform can be expressed as a complex number containing the modulus and argument, and in Appendix B we show the modulus of each of the $12$ pairs of wavenumbers and their phase differences with the meandering streak. Since the summation of these $12$ pairs as shown in figure \ref{fig:stretching_intensity} reflects around $91\%$ of the summation of all modes, we use this summation as a measure of the streamwise vortex stretching effect.


\subsection{Intensity of the characteristic terms of regeneration cycle}


\begin{figure}
\centering
\putfig{}{\includegraphics[width=14.5cm,trim={1cm 1cm 0 0}, clip]{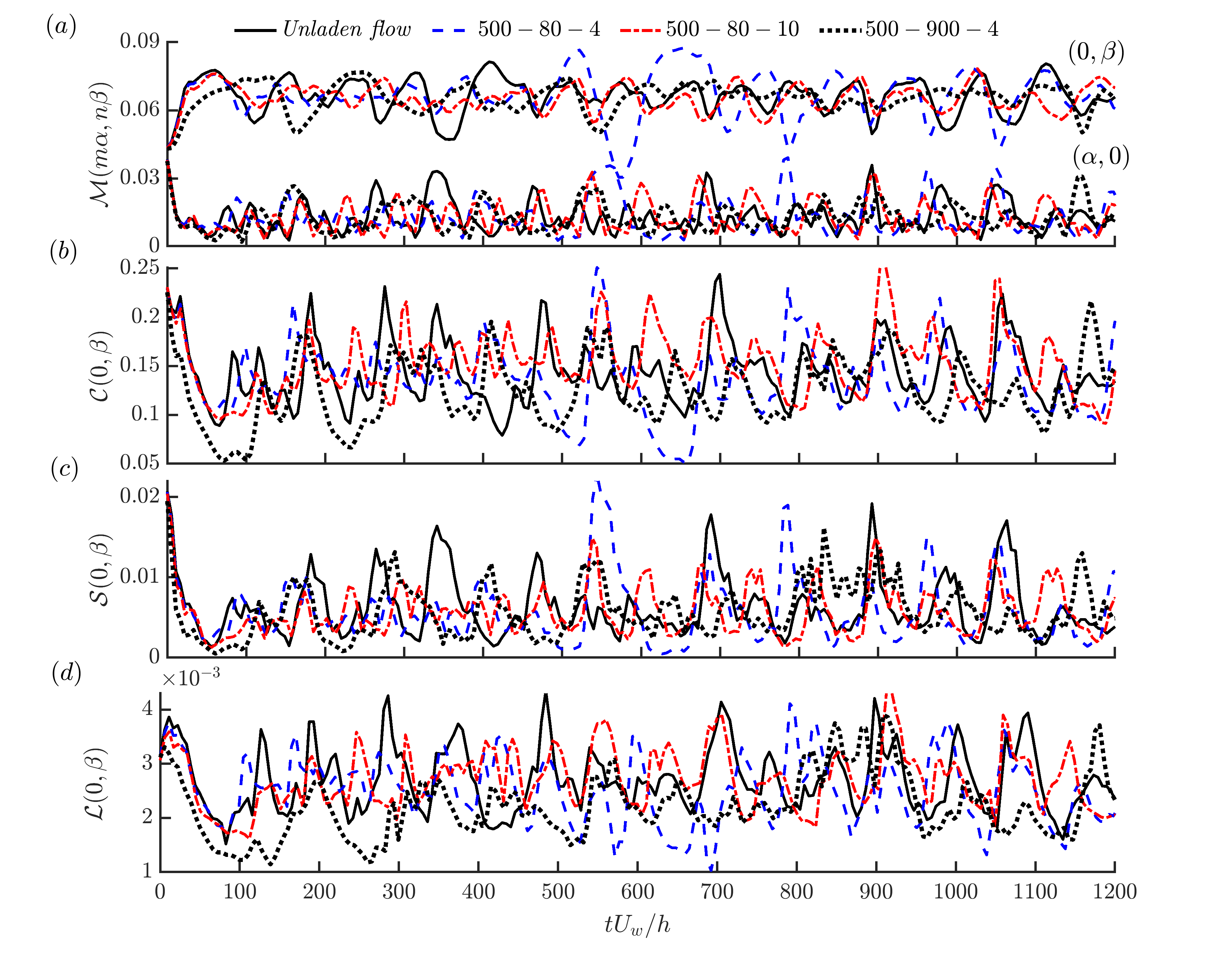}} \quad
  \caption{ Modal decomposition of the various sub-steps composing the regeneration cycle as shown in figure \ref{fig:Figure_regenerationcycle} and outlined in section \ref{sec:regeneration_cycle}. Four cases $500-0-0$ (unladen), $500-80$ ($500-80-4$ and $500-80-10$ with $St_{turb}=0.056$, turbulence enhancement) and $500-900-4$ ($St_{turb}=0.625$, turbulence attenuation) are shown. $(a)$ $\mathcal{M}(n\alpha,m\beta)$ as in equation (\ref{eq:fft_LSS}) representing turbulent kinetic energy contained in the streaks; $(b)$ $\mathcal{C}(0,\beta)$ as in equation (\ref{eq:fft_LSS}) representing for the intensity of the LSV; $(d)$ $\mathcal{C}(0,\beta)$ representing the lift-up effect induced by LSV to enhance LSS as in equation (\ref{eq:fft_Liftup}). The initial $1200$ time units ($h/U_w$) are shown.}
\label{fig:1d-modal}
\end{figure}

\begin{table}
\centering
\def~{\hphantom{0}}
\begin{tabular}{lcccccccccc}

\multicolumn{1}{c}{} & \multicolumn{2}{c}{$\mathcal{M}(0,\beta)$} & \multicolumn{2}{c}{$\mathcal{M}(\alpha,0)$} & \multicolumn{2}{c}{$\mathcal{C}(0,\beta)$} & \multicolumn{2}{c}{$\mathcal{L}(0,\beta)$} & \multicolumn{2}{c}{$\mathcal{S}(0,\beta)$} \\

\\

\multicolumn{1}{c}{} & \multicolumn{10}{c}{$Average~(\times 10^{-2})$}\\

Unladen flow & \multicolumn{2}{c}{6.64} & \multicolumn{2}{c}{1.32} & \multicolumn{2}{c}{14.0} & \multicolumn{2}{c}{0.27} & \multicolumn{2}{c}{0.63} \\

500-80-4 & \multicolumn{2}{c}{6.64} & \multicolumn{2}{c}{1.30} & \multicolumn{2}{c}{14.1} & \multicolumn{2}{c}{0.26} & \multicolumn{2}{c}{0.58} \\

500-80-10 & \multicolumn{2}{c}{6.62} & \multicolumn{2}{c}{1.26} & \multicolumn{2}{c}{14.7} & \multicolumn{2}{c}{0.26} & \multicolumn{2}{c}{0.55} \\


500-900-4 & \multicolumn{2}{c}{6.62} & \multicolumn{2}{c}{1.28} & \multicolumn{2}{c}{12.4} & \multicolumn{2}{c}{0.22} & \multicolumn{2}{c}{0.55} \\

\\

\multicolumn{1}{c}{} & \multicolumn{10}{c}{$Period~(h/U_w)$}\\

Unladen flow & \multicolumn{2}{c}{104} & \multicolumn{2}{c}{104} & \multicolumn{2}{c}{104} & \multicolumn{2}{c}{99} & \multicolumn{2}{c}{104} \\

500-80-4 & \multicolumn{2}{c}{94} & \multicolumn{2}{c}{89} & \multicolumn{2}{c}{99} & \multicolumn{2}{c}{89} & \multicolumn{2}{c}{94} \\

500-80-10 & \multicolumn{2}{c}{89} & \multicolumn{2}{c}{84} & \multicolumn{2}{c}{80} & \multicolumn{2}{c}{84} & \multicolumn{2}{c}{84} \\


500-900-4 & \multicolumn{2}{c}{121} & \multicolumn{2}{c}{121} & \multicolumn{2}{c}{121} & \multicolumn{2}{c}{105} & \multicolumn{2}{c}{116} \\

\end{tabular}
\caption{The average and period of five characteristic terms of the regeneration cycle. The error of the period due to the resolution of the signals is in the range of $\pm 2.5$ time units.}
\label{tab:Table_2}
\end{table}

Figure \ref{fig:1d-modal} shows the temporal evolution of the quantities defined in the previous section: the wall-normal integrated amplitude of LSSs ($\mathcal{M}(0,\beta)$, figure \ref{fig:1d-modal}(a)), the strength of the meandering streak (represented by mode $\mathcal{M}(\alpha,0)$, figure \ref{fig:1d-modal}(a)), LSV strength (represented by $\mathcal{C}(0,\beta)$, figure \ref{fig:1d-modal}(b)), the vortex stretching term (represented by $\mathcal{S}(0,\beta)$ which is the summation of all 12 pairs of $p= \pm 1$ and $r= \mp 1$ in combination with $q=-2$ to $3$ and $s=3$ to $-2$, figure \ref{fig:1d-modal}(c)) and the lift-up term (represented by $\mathcal{L}(0,\beta)$ mode, figure \ref{fig:1d-modal}(d)). Clearly, all of the signals are fluctuating in time with a similar period corresponding the three regeneration steps as shown in figure \ref{fig:Figure_regenerationcycle}. In single-phase flow, \citet{hamilton1995regeneration} estimated this period is slightly less than $100$ time units ($h/U_w$) at $\Rey_b=500$. In flow influenced by low-inertia finite-size particles, \citet{wang2017modulation} also observed a similar period. 

Figure \ref{fig:1d-modal} shows that the amplitudes of $\mathcal{M}(0,\beta)$ and $\mathcal{M}(\alpha,0)$ are diminished by the inertial particles of case $500-900-4$ (dotted black line), which is consistent with the behavior observed by \citet{wang2017modulation} for finite-size particles. However for low particle inertia, the amplitude of $\mathcal{M}(\alpha,0)$ is nearly unchanged due to low mass fraction (case $500-80-4$) whereas it is suppressed with the presence of more particles of the same Stokes number (case $500-80-10$). The presence of a large number of inertial particles (either when they enhance or attenuate the turbulence) tends to stabilize the amplitude of LSSs ($\mathcal{M}(0,\beta)$) but not significantly affect the time-averaged intensity of the streaks. 

The time-averaged intensities of $\mathcal{C}(0,\beta)$, $\mathcal{S}(0,\beta)$, and $\mathcal{L}(0,\beta)$ are provided in table \ref{tab:Table_2}, where we see that these four characteristic terms are all suppressed when turbulence attenuation occurs due to particles (case $500-900-4$). For cases with turbulence enhancement, however (cases $500 - 80-10$ and $500-80-4$), the primary difference is the amplified intensity of the circulation ($\mathcal{C}(0,\beta)$, also seen in figure \ref{fig:1d-modal}(b)), and this amplification increases with particle concentration. The strengthened LSVs are critical for sustaining the turbulence and thus are consistent with the observed lower transitional Reynolds number ($\Rey_c \sim 290$ in case $500-80-4$ versus $\Rey_c \sim 320$ for single phase flow as shown in figure \ref{fig:Transition}(a)). As found by \citet{hamilton1994streamwise}, there is a minimum threshold of circulation below which the turbulence collapses and the flow becomes laminar. At lower Reynolds numbers and/or higher particle concentrations, we indeed see that this is the case (not shown here), and it appears that the primary effect of the particles is to modify the LSVs strength both during stabilization and destabilization.

\subsection{Periodic behavior and phase difference of the regeneration cycle}

To better understand the effects of on the timing of the regeneration cycle, we calculate a temporal auto-correlation of the five signals in figure \ref{fig:1d-modal}:

\begin{equation}\label{eq:autocorrelation}
R_{ss}(\Delta t)=\frac{ \overline{s{^\prime}  (t) ~ s{^\prime} (t+\Delta t)}  }{s{^\prime}_{rms} ^2},
\end{equation}

\noindent where $s{^\prime}$ is the fluctuation of a signal with respect to the time-averaged value.

\begin{figure}
\centering
\putfig{}{\includegraphics[width=14.5cm,trim={2cm 0 0 0}, clip]{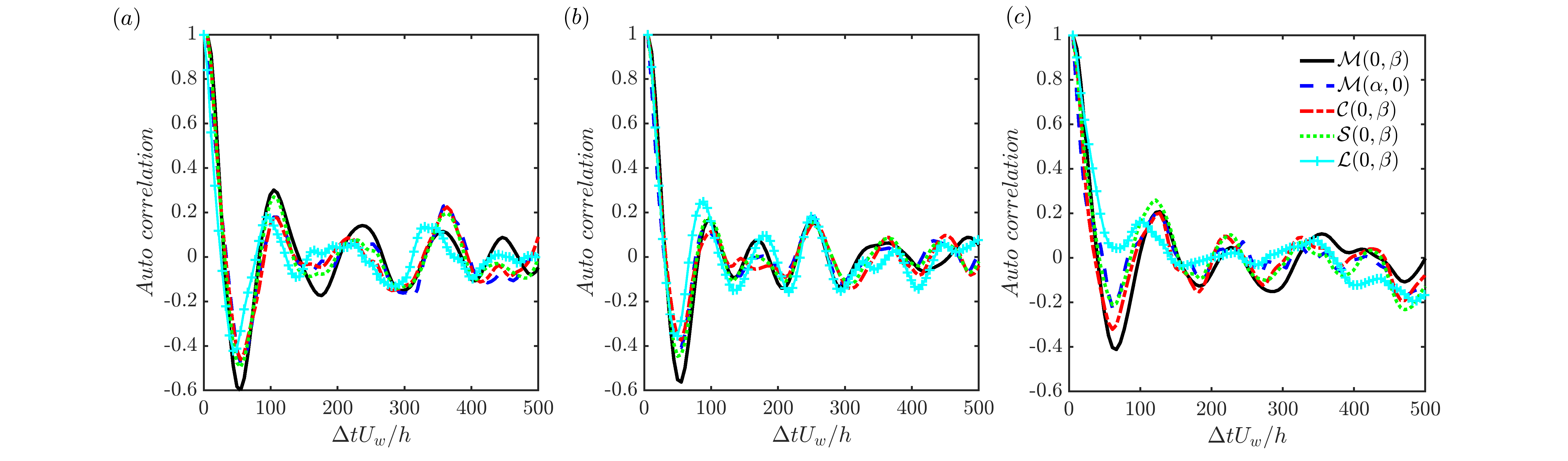}} \quad
  \caption{Temporal auto-correlation functions of the five signals shown in figure \ref{fig:1d-modal}. $(a)$ Single-phase flow, case $500-0-0$; Suspension flow, $(b)$ case $500-80-4$; $(c)$ case $500-900-4$.}
\label{fig:Autocorrelation}
\end{figure}

The temporal auto-correlation is calculated over $3600$ time units to ensure converged statistics. Figure \ref{fig:Autocorrelation}(a) plots the five temporal auto-correlations in single-phase flow, case $500-80-4$ is shown in figure \ref{fig:Autocorrelation}(b), and case $500-900-4$ is shown in figure \ref{fig:Autocorrelation}(c). The time difference between the first two maximums of the correlation coefficient is taken as the period of the cycle and these are summarized in table \ref{tab:Table_2}. We can see that the period changes due to the presence of the particles, and that in cases of turbulence enhancement, the period is shortened, while for turbulence suppression, the period is lengthened. Along these lines, in single-phase flow, \citet{hamilton1995regeneration} observed that the period of the regeneration cycle is shortened at lower Reynolds numbers, e.g. 18 cycles in 1500 time units at $\Rey_b=400$ versus 16 cycles in 1500 time units at $\Rey_b=500$. Therefore comparing to single-phase flow, the period of the regeneration cycle increases in case $500-900-4$ (turbulence is attenuated, behaving as a lower Reynolds in single-phase flow) whereas the period of the regeneration cycle decreases in cases $500-80-4$ and $500-80-10$ (turbulence is enhanced, behaving as a higher Reynolds in single-phase flow).


\begin{figure}
\centering
\putfig{}{\includegraphics[width=14.5cm,trim={2cm 1.5cm 0 0}, clip]{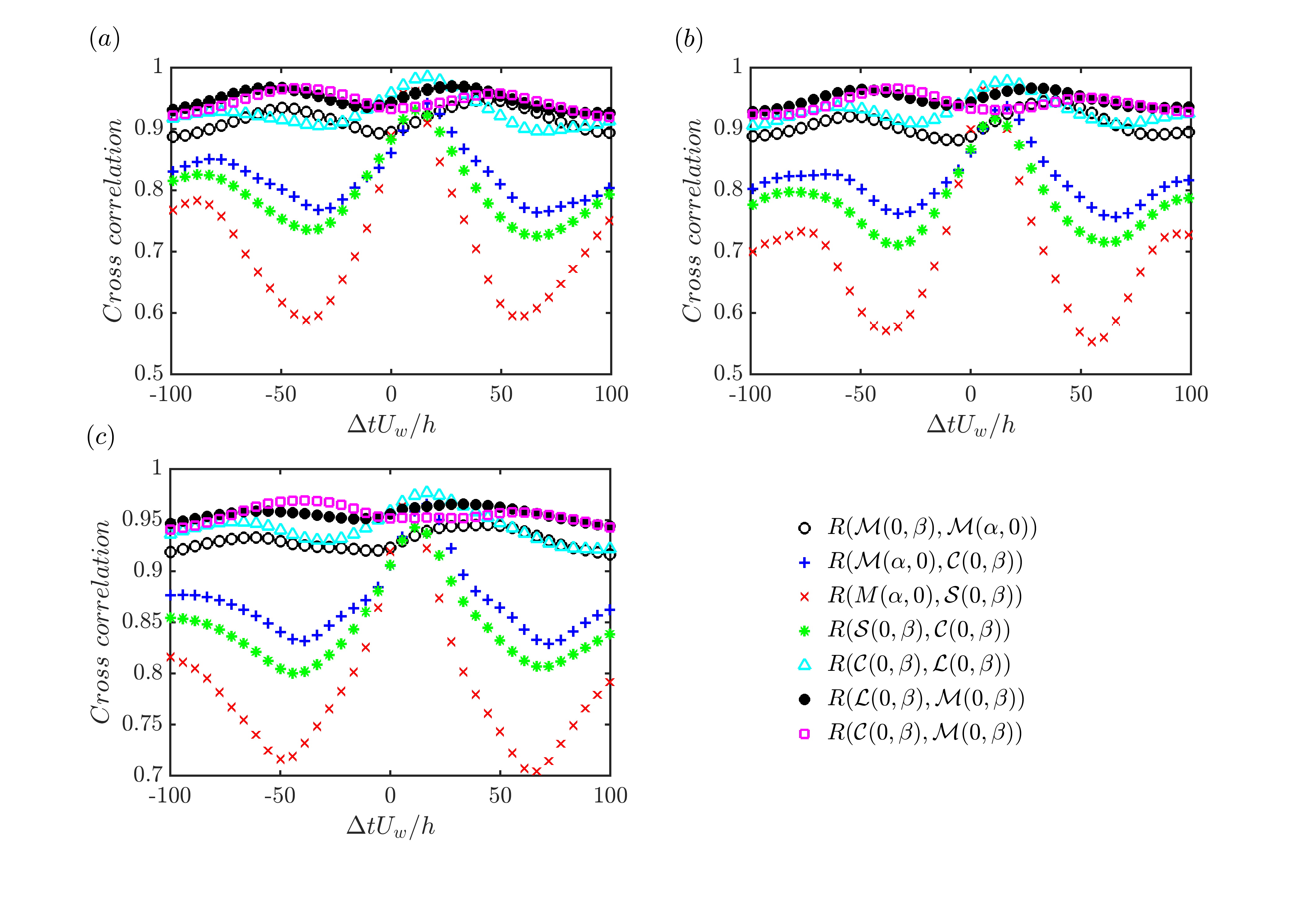}} \quad
  \caption{Temporal cross-correlation functions of the five signals shown in figure \ref{fig:1d-modal}. $(a)$ Single-phase flow, case $500-0-0$; $(b)$ Suspension flow, case $500-900-4$. The corresponding magnified figures with phase shift in the range of $-100$ to $100$ time units are shown in $(c)$ and $(d)$.}
\label{fig:Crosscorrelation}
\end{figure}

The regeneration cycle consists of three sequential sub-processes sketched in figure \ref{fig:Figure_regenerationcycle}: streak formation, streak breakdown and streamwise vortex regeneration. The key mechanisms are: the LSSs are generated by a linear lift-up process by the LSVs, the LSSs break down to meandering streaks by complex nonlinear interactions, and wavy streaks (x-dependent, contributing to ${\partial u} / {\partial x}$) interacts with streamwise vorticity ($\omega_x$) to strengthen the LSVs, which is another nonlinear process. This cycle is recognized as a robust, spatial-temporal evolution and time-stable self-sustaining process. 
We noted above that the period of this cycle (e.g. signal $\mathcal{M}(0,\beta)$) is around $104$ time units in single-phase flow, $94$ time units in case $500-80-4$ and $121$ time units in case $500-900-4$. It is instructive to further analyze the phase difference during the regeneration cycle, especially in order to quantify the inertial particle effect during turbulence modulation. As can be seen in figure \ref{fig:1d-modal}, the LSSs ($\mathcal{M}(0,\beta)$) and the meandering streaks ($\mathcal{M}(\alpha,0)$) nearly have opposite phase. Two subsequent sub-steps, the vortex stretching ($\mathcal{S}(0,\beta)$) and the lift-up ($\mathcal{L}(0,\beta)$) induced by LSVs, are closely synchronous with the temporal evolution of circulation ($\mathcal{C}(0,\beta)$). Vortex stretching, lift-up, and circulation all seem to remain in-phase with the meandering streaks. To verify this, we perform temporal cross-correlation study of the five signals based on equation (\ref{eq:crosscorrelation}). 

\begin{equation}\label{eq:crosscorrelation}
R_{s_1 s_2}(\Delta t)=\frac{ \overline{s_1{^\prime}  (t+\Delta t) ~ s_2{^\prime} (t)}  }{s_1{^\prime}_{rms} s_2{^\prime}_{rms}},
\end{equation}
where $s_1{^\prime}$ and $s_2{^\prime}$ are the fluctuations of signals $s_1$ and $s_2$ with respect to the time-averaged value. For instance, $R(\mathcal{M}(0,\beta), \mathcal{M}(\alpha,0))$ represents temporal cross-correlation between $ \mathcal{M}(0,\beta)$ (LSS) and $\mathcal{M}(\alpha,0)$ (meandering streaks), thereby providing information on temporal lag between the two processes. If $R(\mathcal{M}(0,\beta), \mathcal{M}(\alpha,0))$ is positive, it reflects that $ \mathcal{M}(0,\beta)$ occurs prior in time to $\mathcal{M}(\alpha,0)$ during one regeneration cycle --- the physical explanation for this specific case is that the LSS is formed, and subsequently breaks down into meandering streaks.


\begin{table}
\centering
\begin{tabular}{lccccccc}

~ & $\mathcal{M}(0,\beta),\mathcal{M}(\alpha,0)$ & $\mathcal{M}(\alpha,0),\mathcal{C}$ & $\mathcal{M}(\alpha,0),\mathcal{S}$ & $~\mathcal{S},\mathcal{C}~$ & $~\mathcal{C},\mathcal{L}~$ & $~\mathcal{L},\mathcal{M}~$ & $~\mathcal{C},\mathcal{M}~$ \\

\\

Unladen flow 	& 40 & 17 & 8 & 11 & 17 & 30 & 44 \\
500-80-4 		& 34 & 16 & 6 & 11 & 11 & 30 & 40 \\
500-80-10 		& 28 & 16 & 6 & 11 & 16 & 28 & 34 \\
500-900-4 		& 44 & 17 & 8 & 11 & 17 & 38 & 58 \\

\end{tabular}
\caption{Time difference (time units) between connected sub-steps of the regeneration cycle. The error due to the resolution of the signals is in the range of $\pm 2.5$ time units.}
\label{tab:Table_3}

\end{table}

Figure \ref{fig:Crosscorrelation} illustrates the phase difference between $\mathcal{M}(0,\beta)$, $\mathcal{M}(\alpha,0)$, $\mathcal{C}(0,\beta)$, $\mathcal{S}(0,\beta)$ and $\mathcal{L}(0,\beta)$ averaged over $3600$ time units. Starting from the LSS ($\mathcal{M}(0,\beta)$) stage, we have chosen $7$ correlation coefficients between every pair of connected sub-steps. Figure \ref{fig:Crosscorrelation}(a) shows these coefficients for single-phase flow, in figure \ref{fig:Crosscorrelation}(b) for case $500-80-4$, and in figure \ref{fig:Crosscorrelation}(c) for case $500-900-4$. We quantify the phase shifts in table \ref{tab:Table_3}.
Although inertial particles in case $500-900-4$ have been found to attenuate the turbulence activity, the inertial particles actually do not significantly alter the basic regeneration cycle or its timing. The shortened or lengthened period of the regeneration cycle is mainly due to the particle modulation of streak breakdown from $ \mathcal{M}(0,\beta)$ (LSS) to $\mathcal{M}(\alpha,0)$ (meandering streaks) and streak formation from $ \mathcal{C}(0,\beta)$ (LSV) to $ \mathcal{M}(0,\beta)$ (LSS). Here we should point out that the vortex regeneration is not necessary for every cycle; it might be absent for some cycles \citep{hamilton1995regeneration}, but its minimum value has to be above a threshold to produce unstable streaks.

\begin{figure}
  \centerline{\includegraphics[width=15cm,trim={1cm 0 1cm 0}, clip]{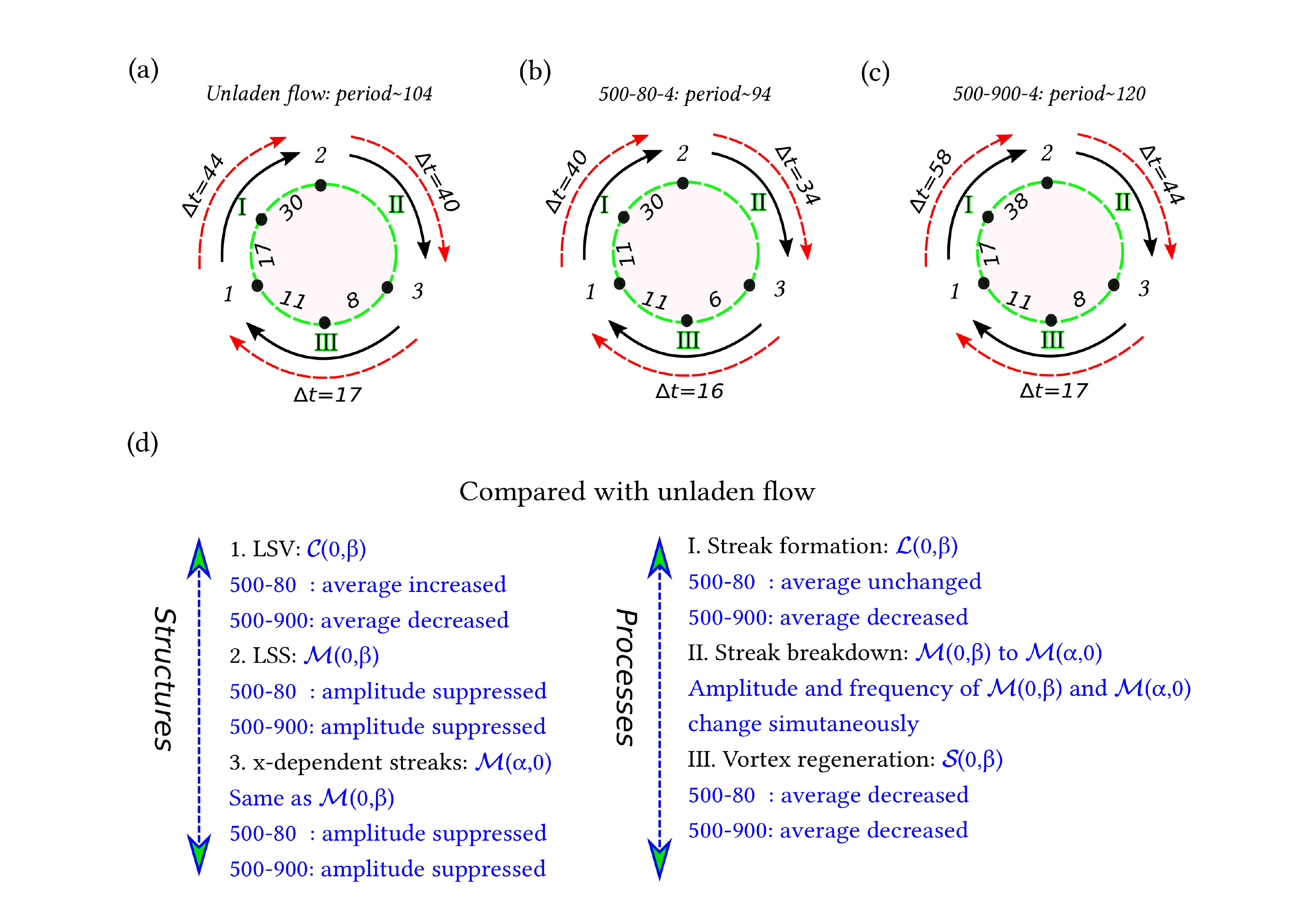}}
  \caption{Sketch of phase difference between connected sub-steps and the period of the regeneration cycle in minimal unit, $(a)$ case $500-0-0$ in comparison with $(b)$ case $500-80-4$ and $(c)$ case $500-900-4$. Qualitative comparison of the averaged value and the amplitude of the signals with single-phase flow are illustrated in the bottom.}
\label{fig:Figure_regeneration_temporal}
\end{figure}

Finally, we can summarize the primary effect of inertial particles on turbulence regeneration by continuing to use case $500-80$ (including $500-80-4$ and $500-80-10$ with $St_{turb}=0.056$) and case $500-900-4$ ($St_{turb}=0.625$) as representative examples of turbulence enhancement and turbulence attenuation, respectively. A temporal evolution of the regeneration cycle can be seen in the schematic sketch in figure \ref{fig:Figure_regeneration_temporal} (a-c). In this figure, we highlight the key turbulent structures and sub-steps by their corresponding metrics. The summation of the transition from $\mathcal{M}(0,\beta)$ to $\mathcal{M}(\alpha,0)$, from $\mathcal{M}(\alpha,0)$ to $\mathcal{C}(0,\beta)$, and from $\mathcal{C}(0,\beta)$ to $\mathcal{M}(0,\beta)$, yields roughly $101$ time units in single-phase flow, $90$ time units in case $500-80-4$ and $119$ time units in case $500-900-4$ corresponding to the whole period $104$ time units, $94$ time units and $120$ time units, respectively. Taking into consideration the resolution of signals, it is reasonable to believe that the summed phase differences are consistent with the calculated period of the regeneration cycle.

The regeneration cycle is therefore a temporal sequence of processes which has a stable periodicity, even under the influence of inertial particles --- phase differences between linked sub-steps are nearly unchanged by the presence of particles except streaks formation and breakdown. Instead, during turbulence enhancement (case $500-80$), as can be seen in figure \ref{fig:Figure_regeneration_temporal}(d), the particles simply strengthen the magnitude of the LSVs ($\mathcal{C}(0,\beta)$), whereas during turbulence attenuation (case $500-900-4$), the particles substantially reduce the magnitude of key turbulent structures and their corresponding sub-steps. Once the magnitude of the LSVs ($\mathcal{C}(0,\beta)$) is below a certain threshold of the magnitude of the LSVs (as in case $380-900-4$), particles act to abruptly shut off the regeneration, thereby delaying turbulent-to-laminar transition. It is worth noting that a similar upward shift of the transitional Reynolds owing to the vortex suppression is also observed during turbulence modulation laden with polymer additives --- the modifications to the `exact coherent states' are due to the suppression of the streamwise vortices due to the polymers exert an opposite force to the fluid motion in the vortices and weaken the streamwise vortices \citep[see][]{stone2004polymer}. \\

\section{Concluding remarks}\label{sec:Conclusion}

This work investigates the effect of inertial pointwise particles on the self-sustained process of the coherent structures in turbulent plane Couette flow. A `miniunit' configuration with a low Reynolds number slightly above the onset of transition not only highlights the key turbulent structures making up the regeneration cycle \citep[the miniunit and regeneration cycle are interpreted by][]{hamilton1995regeneration}, but can be simulated with much reduced computational cost. Two-way coupling of Lagrangian point particles with direct numerical simulations of Eulerian flow is the methodology used in this work.

Through examining the particle inertial effect on transition, we find that lower particle inertia tends to advance the transition from laminar to turbulent flow whereas the high particle inertia tends to delay the transition. These two limiting cases reflect the competition between the destabilization of the extra mixed-phase density and the stabilization due to the extra dissipation caused by the particle drag. While the nonlinear simulations capture features such as preferential accumulation, these results are in qualitative agreement with the linear stability analysis of \citet{saffman1962stability}, who assumed uniformly distributed particles. In agreement with other studies of particle-laden, wall-bounded turbulent flow, the particle response time alters where they tend to collect. The particle spatial distribution in turbulence is found to relate with the particle response time scaled by the characteristic LSV turnover time, which in turn affects the regeneration cycle. In the present domain, particles can either reside in the so-called LSS regions (high strain) or in the LSV regions (high vorticity). Low particle inertia (e.g. $St_{turb}=0.056$) leads to long residence times in the same LSV, whereas large particle inertia (e.g. $St_{turb}=0.625$) results in particles crossing the LSVs leading to a more homogeneous particle distribution. Between these extremes, particles with a moderate response time scale (e.g. $St_{turb}=0.347$) are expelled from the LSVs and then collect in the LSSs leading to more particles accumulate in the near wall region.

The particulate turbulent kinetic energy is highly related to their spatial distribution and the associated slip velocity. Due to the preferential distribution of moderate inertia particles (e.g. $St_{turb}=0.347$) in the LSSs, the particulate turbulent kinetic energy is higher than the fluid's in the streamwise direction whereas it is lower in wall-normal and spanwise directions. Particles with a large response time scale (e.g. $St_{turb}=0.625$) cannot follow the streamlines, resulting in a more homogeneous distribution, which results in a more homogeneous turbulent kinetic energy of the particulate phase, however it still exceeds the fluid phase kinetic energy in the steamwise direction.


Modal analysis is performed to examine the effect of the particles on the nonlinear regeneration cycle. For a representative set of cases at $\Rey_b = 500$, we quantitatively obtain its period to be roughly $104$ time units in single-phase flow (case $500-0-0$), whereas the period is around $94$ time units in case $500-80-4$ (turbulence enhancement happens in this case with $St_{turb}=0.056$), and the period is around $121$ time units in case $500-900-4$ (turbulence attenuation happens in this case with $St_{turb}=0.625$). Furthermore, the phase differences between the linked sub-steps are obtained based on the maximum cross-correlation coefficient. The duration of each sub-step of the cycle as a fraction of the whole period of the regeneration cycle is about $43\%$ (resp. $44\%$ and $48\%$) from LSV to LSS, $40\%$ (resp. $38\%$ and $38\%$) from LSS to meandering streaks, and $17\%$ (resp. $18\%$ and $14\%$) from meandering streaks to LSV, in single-phase flow (resp. $500-80-4$ and $500-900-4$). The vortex regeneration process (from meandering streaks to LSV) is very fast whereas the lift-up process and streaks breakdown process are relatively slow. The low inertial particles (e.g. $St_{turb}=0.056$) enhance the turbulence which shortens the period of the regeneration cycle whereas the high inertial particles (e.g. $St_{turb}=0.625$) attenuate the turbulence which lengthens the period of the regeneration cycle.

Even though inertial particles are found to have little effect on the phase differences between the linked sub-steps compared with the whole period, the intensity of circulation, lift-up, and steramwise vorticity stretching are greatly suppressed (the reductions are about $15\%$ compared to single-phase flow in case $500-900-4$) due to the presence of heavy inertial particles (e.g. $St_{turb}=0.625$). The reduced intensity of circulation does not directly change the mean value of the streaks, while the amplitude of the LSSs is stabilized significantly due to the suppression of the LSVs following a suppressed lift-up effect. The simulations show that it is streamwise particle-fluid coupling in an annulus region in the outer region of the LSVs coinciding with the region where streamwise vortex stretching is most active which plays the key role in suppressing the streamwise vortex stretching term and further deceasing the LSVs. However during turbulence enhancement in cases $500-80$, the circulation of LSVs is strengthened so that the minimum circulation threshold happens at a lower transition Reynolds number.

\section{Acknowledgement}\label{sec:Acknowledgement}
The authors acknowledge grant G00003613-ArmyW911NF-17-0366 from the U.S. Army Research Office for financial support. Computational resources were provided by the Notre Dame Center for Research Computing. G. W is grateful to Professor M. Abbas of Laboratoire de G\'{e}nie Chimique (LGC) and Professor E. Climent of Institut de M\'{e}canique des Fluides de Toulouse (IMFT) for very valuable discussions and constructive suggestions regarding the turbulent regeneration cycle and particle suspension flow.\\

\section*{Appendix A. Effect of particle response time in Taylor-Green vortex}\label{sec:TGV}

In turbulent pCf, the counter-rotating LSVs fill the Couette gap and the accompanying LSSs occur near the corners of each LSV. These streamwise independent structures can be viewed as a three-dimensional flow with a single Fourier mode in spanwise direction. As an idealized model of this structure, the undissipated steady Taylor-Green vortex flow in the domain $0 \leqslant x, y \leqslant 2\pi$ has two pairs of counter-rotating vortices and a strong strain rate region near the corners of each cell, and is therefore a simple but robust analytical framework for the study of particle distribution in the LSVs and LSSs since the Taylor-Green vortex can capture key features of particle dispersion in vortical or high strain rate regions \citep[see][]{maxey1987motion}.

In section \ref{sec:Configuration}, we define a Stokes number $St_{turb}$ that is the ratio between particle response time with the LSV turnover time scale. Similarly, we here define a Stokes number $St_{TGV}$ which is the ratio between particle response time with the TGV turnover time scale. \citet{massot2007eulerian} proposed a threshold for $St_{TGV}$ as $1/(8\pi)$, below which the particles remain forever inside the TGV without crossing trajectories from one vortex to another.

\begin{figure}
\centering
\putfig{}{\includegraphics[width=14.5cm,trim={4.5cm 2cm 0 2cm}, clip]{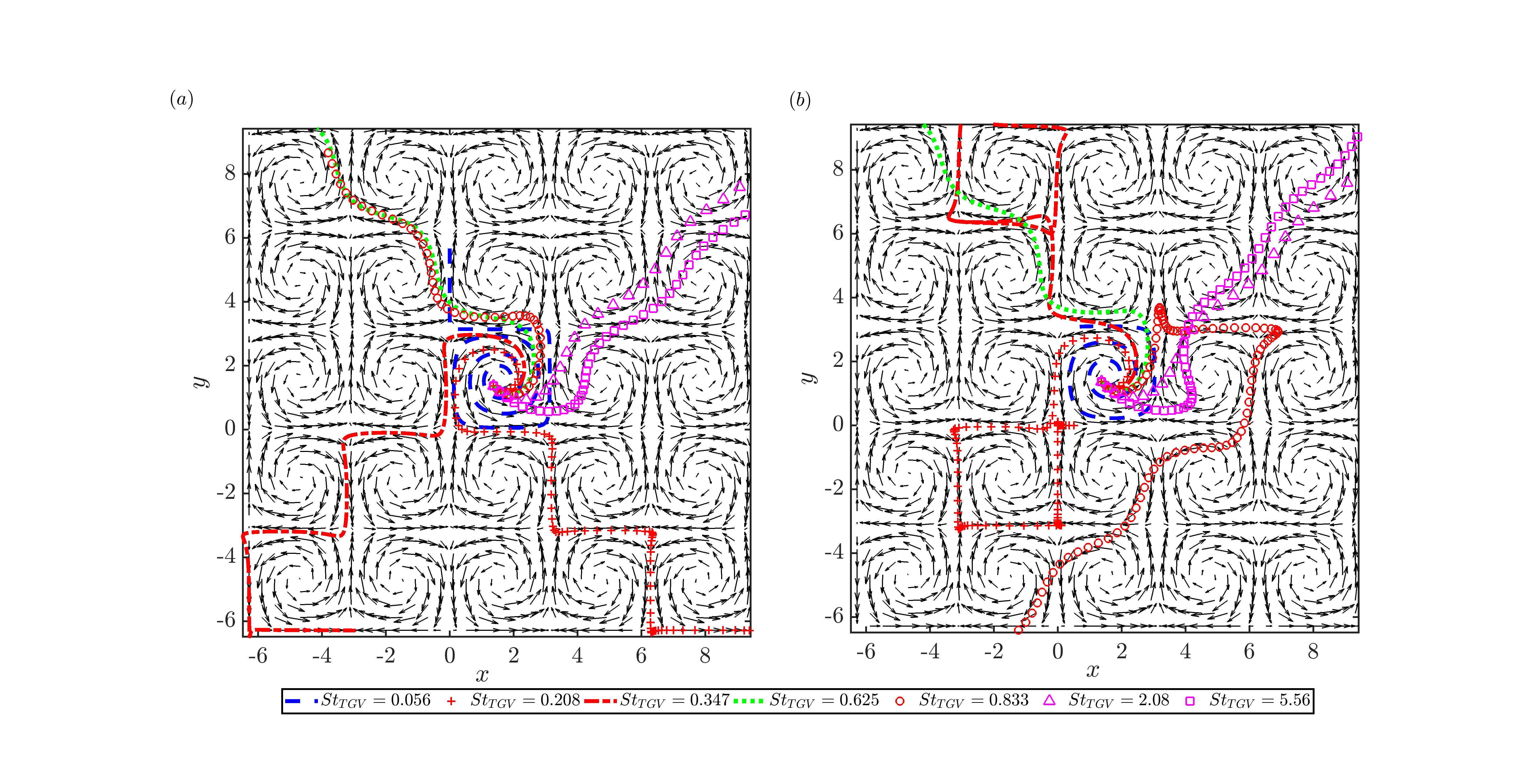}} \quad
  \caption{Inertial particle trajectories released initially at $(1.73,1.73)$ with different $St_{TGV}$ in Taylor-Green vortex. $(a)$ Maxey-Riley equation with lift-force; $(b)$ the same condition as $(a)$ without Basset history force.}
\label{fig:TGV}
\end{figure}

The trajectory of a single heavy particle in TGV is numerically predicted by the Maxey-Riley equation \citep{maxey1983equation} for point particles at low Reynolds number. The numerical method has been validated by comparing with \citet{daitche2013advection}. Figure \ref{fig:TGV}(a) shows an inertial particle released at $x=1.37,~y=1.37$ and solved with one-way coupling (eliminating only the Basset history force is shown in figure \ref{fig:TGV}(b)). Comparing figures \ref{fig:TGV}(a) with (b), the existence of the history force tends to keep the particles inside the vortical structures and reduce the slip velocity, and its inclusion is more important at lower Stokes number. The Stokes number used in this simulation ($St_{TGV}$) is consistent with our turbulent pCf ($St_{turb}$). Generally, the particle trajectory can be divided into three categories based on the $St_{TGV}$:\\

At small $St_{TGV}$ (e.g. $St_{TGV}=0.056$ corresponding to $r=80$ in turbulent pCf), the particle prefers to stay inside one cell and then moves along the edges of the cells (high strain rate region) once it moves out of the vortices;

At moderate $St_{TGV}$ (e.g. $St_{TGV}=0.347$ corresponding to $r=500$ in turbulent pCf), the particle residence time inside a cell is decreased with a increase Stokes number, but the particle still follows the edge of the cell (i.e. the high-strain region) after it has been expelled out of the vortices;

At high $St_{TGV}$ (e.g. $St_{TGV}=2.08$ corresponding to $r=3000$ in turbulent pCf), the particle might move across the high vorticity cell, and no longer follows the local streamline due to its high inertia (high local slip velocity due to the unmatched particle response time scale with the eddy turnover time scale).\\

The particle behavior in TGV is analogous to particles distribution in turbulent plane Couette flow with the same particle time scale as shown in figure \ref{fig:Dispersion}(a ,b). The case $500-80-4$ with low-inertia particles ($St_{turb}=0.056$) causing the particles to stay long in single LSV whereas case $500-8000-4$(1way) with high-inertial particles ($St_{turb} = 5.56$) leads to particles moving cross different LSVs once being ejected by the LSSs. Particles with intermediate Stokes numbers, as in case $500-500-4$ with particles $St_{turb} = 0.35$, are expelled from the LSVs and become trapped inside the LSSs.

\section*{Appendix B. Particle effects on $12$ modes of vorticity stretching term}\label{sec:Stretching_pqrs}

The streamwise vorticity transport equation for an incompressible inviscid flow is expressed in equation (\ref{eq:vorticity_dynamic}). The vorticity stretching $+\omega_x \partial u / \partial x$ represents the enhancement of vorticity and is responsible for the cascade process in turbulence. \citet{sendstad1992near} found that the tilting term $-\frac{\partial w}{\partial x} \frac{\partial u}{\partial y}$ is the largest contribution to equation (\ref{eq:vorticity_dynamic}) close to the wall, and \citet{hamilton1995regeneration} showed that the advection (redistribution) and stretching terms are responsible for the newly produced streamwise vorticity needed for LSV regeneration. In the work of \citep{schoppa2002coherent}, they further qualitatively stated that the vortex formation is dominated by stretching of streamwise vorticity, because the spatial and temporal streamwise vortex collapse is in the right places with the streamwise vorticity stretching, coincidently.

In the modal analysis of the regeneration cycle, the generation of streamwise vorticity is mainly caused by the streamwise vortex stretching term $ \widehat{\omega}_x (p\alpha,q\beta) {\partial  \widehat{u}_x (r\alpha,s\beta)/ \partial x}$ in Fourier space, where $p,q,r,s$ must satisfy $p+r=m$ and $q+s=n$. The maximum circulation is always due to the $x$-independent streamwise vorticity with mode $(0,\beta)$ based on our observations. Therefore even though it is only the relations $p+r=0$ and $q+s=1$ which contribute to $\partial \widehat{\omega}_x (0,\beta)/ \partial t$, there are still ${(N_x-1)(N_z-2)}$ pairs of $\widehat{\omega}_x (p\alpha,q\beta)$ and ${\partial  \widehat{u}_x (r\alpha,s\beta)/ \partial x}$ to be examined. In streamwise direction, the work of \citet{hamilton1995regeneration} has simplified above relation to $\alpha$-modes ($p= \pm 1$ and $r= \mp 1$ satisfy $m=p+r=0$) dominant relation where the higher x-wavenumber modes ($\ \vert p \vert > 1$ and $\vert r \vert > 1$ satisfy $m=p+r=0$) contribution to $\partial \widehat{\omega}_x (0,\beta)/ \partial t$ are negligible. Furthermore, we have compared all $(N_z-2)$ spanwise combinations satisfying $n=q+s=1$ and then find the main contribution in spanwise direction as $q=-2$ to $3$ combining with $s=3$ to $-2$ satisfying relation $q+s=1$, which are used to represent the vorticity stretching term in this work (total $12$ pairs of $x$ and $z$ wavenumber).

\begin{figure}
\centering
\putfig{}{\includegraphics[width=12 cm]{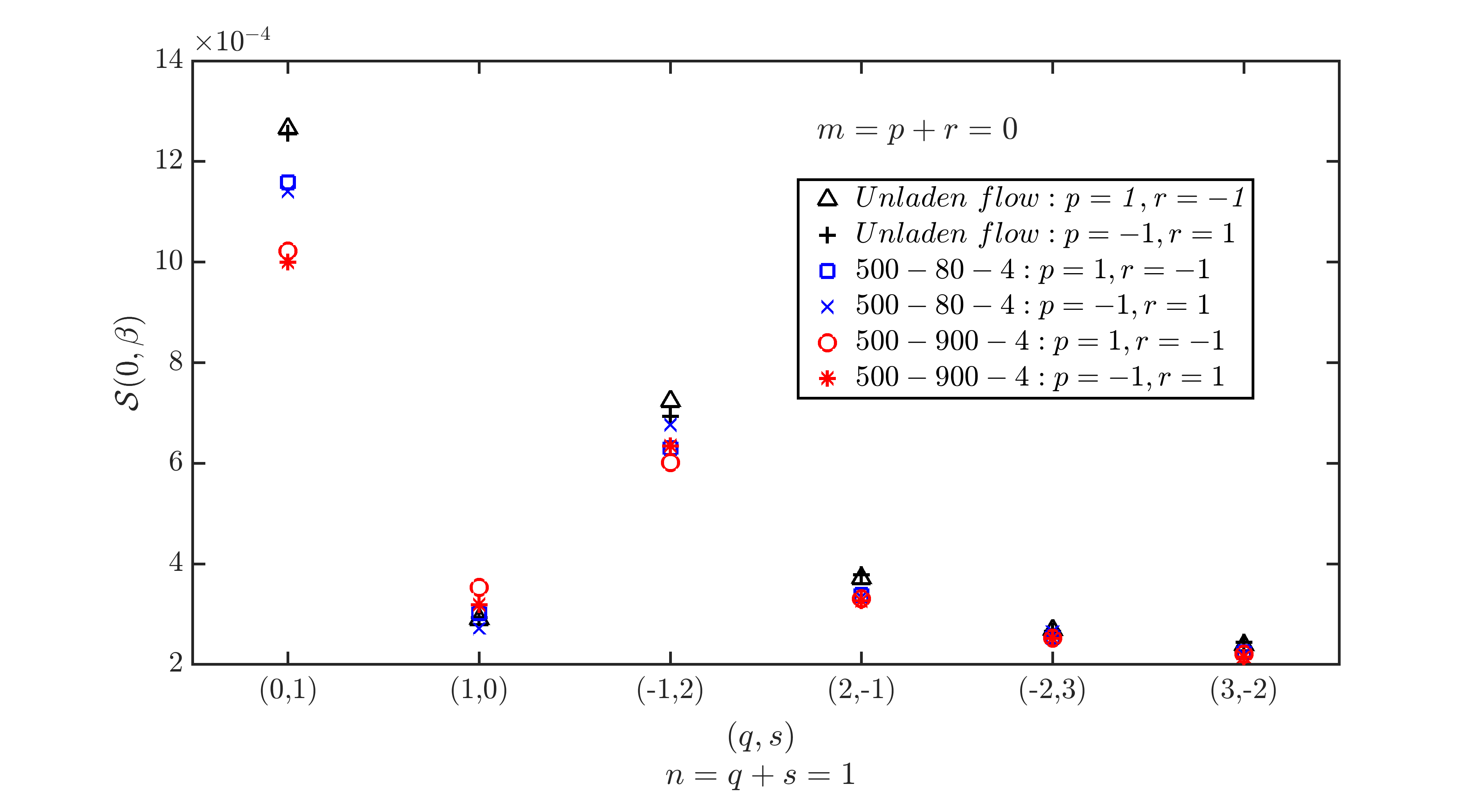}} \quad
  \caption{Temporal average intensity of vorticity stretching term in equation (\ref{eq:fft_stretching}) of 12 pairs of modes. In comparison between single-phase flow ($500-0-0$) with suspension flow ($500-80-4$ and $500-900-4$).}
\label{fig:stretching_intensity}
\end{figure}

Figure \ref{fig:stretching_intensity} plots the wall-normal integrated modulus of vorticity stretching in equation (\ref{eq:fft_stretching}) for the $12$ modes for cases $500-0-0$, cases $500-80-4$ and $500-900-4$. We can see for the same $z$-wavenumber, the integrated modulus of vorticity stretching is similar between $p=1,r=-1$ and $p=-1,r=1$. On the other hand, the contribution from modes $q=0,s=1$ and $q=-1,s=2$ reflects $64 \%$, $61 \%$ and $58 \%$ of the total modulus of all wavenumbers in single-phase flow, case $500-80-4$ and case $500-900-4$, respectively.

\begin{figure}
\centering
\putfig{}{\includegraphics[width=14.5cm,trim={2cm 0 0 0}, clip]{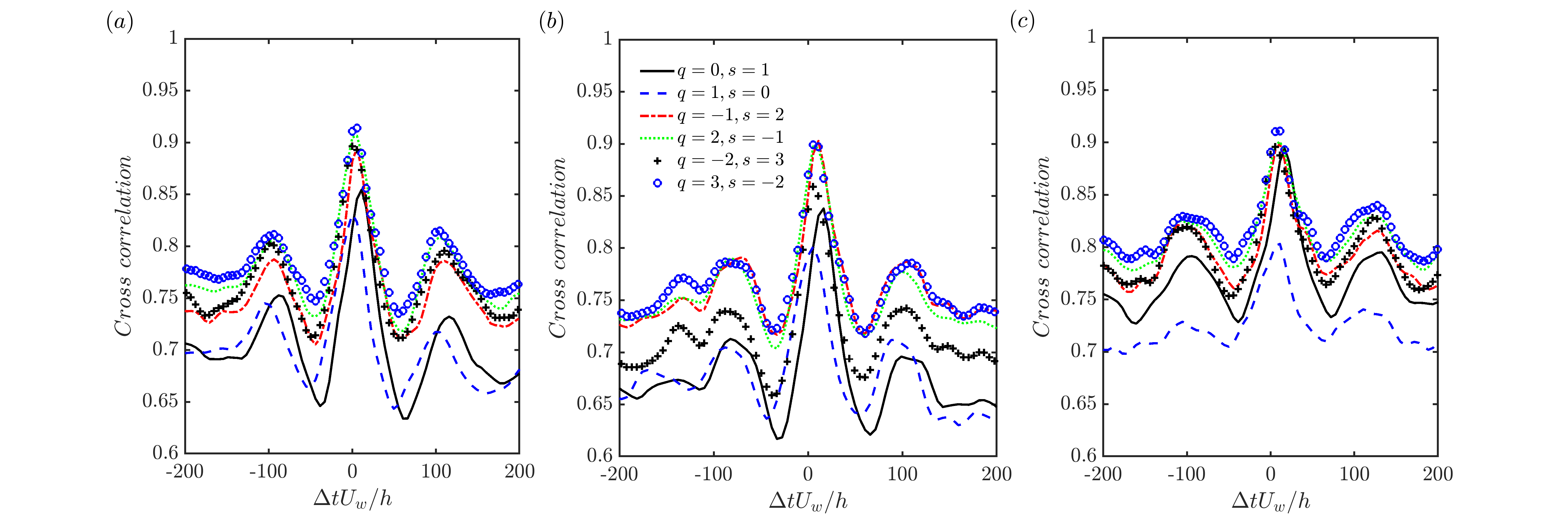}} \quad
  \caption{Phase shifts between vorticity stretching term ($6$ pairs of modes with fixed $p=1,r=-1$) with the meander streak $\mathcal{M}(\alpha,0)$. $(a)$ Single-phase flow ($500-0-0$); Particle-laden flow $(b)$ case $500-80-4$ and $(c)$ case $500-900-4$.}
\label{fig:Crosscorrelation_stretching}
\end{figure}

Figure \ref{fig:Crosscorrelation_stretching} shows the phase difference between $\mathcal{M}(\alpha,0)$ (meandering streaks contributing to $\partial u / \partial x$) with $6$ pairs of mode $q,s$ with fixed $p=1,r=-1$ (the other $6$ pairs of mode $q,s$ with fixed $p=-1,r=1$ give similar results). Except when the phase difference is about $8$ time units for $q=0,s=1$, all others are in the range of $0-5$ time units. We do the summation of all these signals (both modulus and argument angle) to produce figure \ref{fig:Crosscorrelation}.


\bibliographystyle{jfm}
\bibliography{postdoc_ND,BOOKS_JAB,SINGLE-PHASE,TWO-PHASE_JAB}

\begin{thebibliography}{58}
\expandafter\ifx\csname natexlab\endcsname\relax\def\natexlab#1{#1}\fi
\def\au#1{#1} \def\ed#1{#1} \def\yr#1{#1}\def\at#1{#1}\def\jt#1{\textit{#1}}
  \def\bt#1{#1}\def\bvol#1{\textbf{#1}} \def\vol#1{#1} \def\pg#1{#1}
  \def\publ#1{#1}\def\arxiv#1{#1}\def\org#1{#1}\def\st#1{\textit{#1}}

\bibitem[Balachandar \& Eaton(2010)]{balachandar2010turbulent}
{\sc \au{Balachandar, S.} \& \au{Eaton, J.~K.}} \yr{2010}  \at{Turbulent
  dispersed multiphase flow}.  \jt{Annu. Rev. Fluid. Mech.}  \bvol{42},
  \pg{111--133}.

\bibitem[Bech {\em et~al.\/}(1995)Bech, Tillmark, Alfredsson \&
  Andersson]{bech1995investigation}
{\sc \au{Bech, K.~H.}, \au{Tillmark, N.}, \au{Alfredsson, P.~H.} \&
  \au{Andersson, H.~I.}} \yr{1995}  \at{An investigation of turbulent plane
  couette flow at low reynolds numbers}.  \jt{J. Fluid Mech.}  \bvol{286},
  \pg{291--326}.

\bibitem[Brandt(2014)]{brandt2014lift}
{\sc \au{Brandt, L.}} \yr{2014}  \at{The lift-up effect: The linear mechanism
  behind transition and turbulence in shear flows}.  \jt{Eur. J. Mech. B.
  Fluids}  \bvol{47},  \pg{80--96}.

\bibitem[Burton \& Eaton(2005)]{burton2005fully}
{\sc \au{Burton, T.~M.} \& \au{Eaton, J.~K.}} \yr{2005}  \at{Fully resolved
  simulations of particle-turbulence interaction}.  \jt{J. Fluid Mech.}
  \bvol{545},  \pg{67--111}.

\bibitem[Capecelatro {\em et~al.\/}(2018)Capecelatro, Desjardins \&
  Fox]{capecelatro2018transition}
{\sc \au{Capecelatro, J.}, \au{Desjardins, O.} \& \au{Fox, R.~O.}} \yr{2018}
  \at{On the transition between turbulence regimes in particle-laden channel
  flows}.  \jt{Journal of Fluid Mechanics}  \bvol{845},  \pg{499--519}.

\bibitem[Daitche(2013)]{daitche2013advection}
{\sc \au{Daitche, A.}} \yr{2013}  \at{Advection of inertial particles in the
  presence of the history force: Higher order numerical schemes}.  \jt{J.
  Comput. Phys.}  \bvol{254},  \pg{93--106}.

\bibitem[DeSpirito \& Wang(2001)]{despirito2001linear}
{\sc \au{DeSpirito, J.} \& \au{Wang, L.-P.}} \yr{2001}  \at{Linear instability
  of two-way coupled particle-laden jet}.  \jt{Int. J. Multiphase Flow}
  \bvol{27}~(7),  \pg{1179--1198}.

\bibitem[Dimas \& Kiger(1998)]{dimas1998linear}
{\sc \au{Dimas, A.} \& \au{Kiger, K.}} \yr{1998}  \at{Linear instability of a
  particle-laden mixing layer with a dynamic dispersed phase}.  \jt{Phys.
  Fluids}  \bvol{10}~(10),  \pg{2539--2557}.

\bibitem[Dritselis \& Vlachos(2008)]{dritselis2008numerical}
{\sc \au{Dritselis, C.~D.} \& \au{Vlachos, N.~S.}} \yr{2008}  \at{Numerical
  study of educed coherent structures in the near-wall region of a
  particle-laden channel flow}.  \jt{Phys. Fluids}  \bvol{20}~(5),
  \pg{055103}.

\bibitem[Druzhinin \& Elghobashi(1998)]{druzhinin1998direct}
{\sc \au{Druzhinin, O.} \& \au{Elghobashi, S.}} \yr{1998}  \at{Direct numerical
  simulations of bubble-laden turbulent flows using the two-fluid formulation}.
   \jt{Phys. Fluids}  \bvol{10}~(3),  \pg{685--697}.

\bibitem[Eaton \& Fessler(1994)]{eaton1994preferential}
{\sc \au{Eaton, J.~K.} \& \au{Fessler, J.}} \yr{1994}  \at{Preferential
  concentration of particles by turbulence}.  \jt{Int. J. Multiphase Flow}
  \bvol{20},  \pg{169--209}.

\bibitem[Elghobashi(1994)]{elghobashi1994predicting}
{\sc \au{Elghobashi, S.}} \yr{1994}  \at{On predicting particle-laden turbulent
  flows}.  \jt{Appl. Sci. Res.}  \bvol{52}~(4),  \pg{309--329}.

\bibitem[Elghobashi \& Truesdell(1993)]{elghobashi1993two}
{\sc \au{Elghobashi, S.} \& \au{Truesdell, G.}} \yr{1993}  \at{On the two-way
  interaction between homogeneous turbulence and dispersed solid particles. i:
  Turbulence modification}.  \jt{Phys. Fluids A}  \bvol{5}~(7),
  \pg{1790--1801}.

\bibitem[Ellingsen \& Palm(1975)]{ellingsen1975stability}
{\sc \au{Ellingsen, T.} \& \au{Palm, E.}} \yr{1975}  \at{Stability of linear
  flow}.  \jt{Phys. Fluids}  \bvol{18}~(4),  \pg{487--488}.

\bibitem[Garratt(1994)]{garratt1994atmospheric}
{\sc \au{Garratt, J.~R.}} \yr{1994}  \at{The atmospheric boundary layer}.
  \jt{Earth-Science Reviews}  \bvol{37}~(1-2),  \pg{89--134}.

\bibitem[Gore \& Crowe(1989)]{gore1989effect}
{\sc \au{Gore, R.} \& \au{Crowe, C.~T.}} \yr{1989}  \at{Effect of particle size
  on modulating turbulent intensity}.  \jt{Int. J. Multiphase Flow}
  \bvol{15}~(2),  \pg{279--285}.

\bibitem[Hamilton \& Abernathy(1994)]{hamilton1994streamwise}
{\sc \au{Hamilton, J.~M.} \& \au{Abernathy, F.~H.}} \yr{1994}  \at{Streamwise
  vortices and transition to turbulence}.  \jt{J. Fluid Mech.}  \bvol{264},
  \pg{185--212}.

\bibitem[Hamilton {\em et~al.\/}(1995)Hamilton, Kim \&
  Waleffe]{hamilton1995regeneration}
{\sc \au{Hamilton, J.~M.}, \au{Kim, J.} \& \au{Waleffe, F.}} \yr{1995}
  \at{Regeneration mechanisms of near-wall turbulence structures}.  \jt{J.
  Fluid Mech.}  \bvol{287},  \pg{317--348}.

\bibitem[Inoue {\em et~al.\/}(2012)Inoue, Mathis, Marusic \&
  Pullin]{inoue2012inner}
{\sc \au{Inoue, M.}, \au{Mathis, R.}, \au{Marusic, I.} \& \au{Pullin, D.}}
  \yr{2012}  \at{Inner-layer intensities for the flat-plate turbulent boundary
  layer combining a predictive wall-model with large-eddy simulations}.
  \jt{Phys. Fluids}  \bvol{24}~(7),  \pg{075102}.

\bibitem[Jim{\'e}nez(2018)]{jimenez2018coherent}
{\sc \au{Jim{\'e}nez, J.}} \yr{2018}  \at{Coherent structures in wall-bounded
  turbulence}.  \jt{J. Fluid Mech.}  \bvol{842}.

\bibitem[Jim{\'e}nez \& Moin(1991)]{jimenez1991minimal}
{\sc \au{Jim{\'e}nez, J.} \& \au{Moin, P.}} \yr{1991}  \at{The minimal flow
  unit in near-wall turbulence}.  \jt{J. Fluid Mech.}  \bvol{225},
  \pg{213--240}.

\bibitem[Jim{\'e}nez \& Pinelli(1999)]{jimenez1999autonomous}
{\sc \au{Jim{\'e}nez, J.} \& \au{Pinelli, A.}} \yr{1999}  \at{The autonomous
  cycle of near-wall turbulence}.  \jt{J. Fluid Mech.}  \bvol{389},
  \pg{335--359}.

\bibitem[Kawahara \& Kida(2001)]{kawahara2001periodic}
{\sc \au{Kawahara, G.} \& \au{Kida, S.}} \yr{2001}  \at{Periodic motion
  embedded in plane couette turbulence: regeneration cycle and burst}.  \jt{J.
  Fluid Mech.}  \bvol{449},  \pg{291--300}.

\bibitem[Kitoh {\em et~al.\/}(2005)Kitoh, Nakabyashi \&
  Nishimura]{kitoh2005experimental}
{\sc \au{Kitoh, O.}, \au{Nakabyashi, K.} \& \au{Nishimura, F.}} \yr{2005}
  \at{Experimental study on mean velocity and turbulence characteristics of
  plane couette flow: low-reynolds-number effects and large longitudinal
  vortical structure}.  \jt{Journal of Fluid Mechanics}  \bvol{539},
  \pg{199--227}.

\bibitem[Klinkenberg {\em et~al.\/}(2011)Klinkenberg, De~Lange \&
  Brandt]{klinkenberg2011modal}
{\sc \au{Klinkenberg, J.}, \au{De~Lange, H.} \& \au{Brandt, L.}} \yr{2011}
  \at{Modal and non-modal stability of particle-laden channel flow}.  \jt{Phys.
  Fluids}  \bvol{23}~(6),  \pg{064110}.

\bibitem[Klinkenberg {\em et~al.\/}(2014)Klinkenberg, de~Lange \&
  Brandt]{klinkenberg2014linear}
{\sc \au{Klinkenberg, J.}, \au{de~Lange, H.} \& \au{Brandt, L.}} \yr{2014}
  \at{Linear stability of particle laden flows: the influence of added mass,
  fluid acceleration and basset history force}.  \jt{Meccanica}  \bvol{49}~(4),
   \pg{811--827}.

\bibitem[Klinkenberg {\em et~al.\/}(2013)Klinkenberg, Sardina, De~Lange \&
  Brandt]{klinkenberg2013numerical}
{\sc \au{Klinkenberg, J.}, \au{Sardina, G.}, \au{De~Lange, H.} \& \au{Brandt,
  L.}} \yr{2013}  \at{Numerical study of laminar-turbulent transition in
  particle-laden channel flow}.  \jt{Phys. Rev. E}  \bvol{87}~(4),
  \pg{043011}.

\bibitem[Komminaho {\em et~al.\/}(1996)Komminaho, Lundbladh \&
  Johansson]{komminaho1996very}
{\sc \au{Komminaho, J.}, \au{Lundbladh, A.} \& \au{Johansson, A.~V.}} \yr{1996}
   \at{Very large structures in plane turbulent couette flow}.  \jt{J. Fluid
  Mech.}  \bvol{320},  \pg{259--285}.

\bibitem[Lee \& Lee(2015)]{lee2015modification}
{\sc \au{Lee, J.} \& \au{Lee, C.}} \yr{2015}  \at{Modification of
  particle-laden near-wall turbulence: Effect of stokes number}.  \jt{Phys.
  Fluids}  \bvol{27}~(2),  \pg{023303}.

\bibitem[Li {\em et~al.\/}(2001)Li, McLaughlin, Kontomaris \&
  Portela]{li2001numerical}
{\sc \au{Li, Y.}, \au{McLaughlin, J.~B.}, \au{Kontomaris, K.} \& \au{Portela,
  L.}} \yr{2001}  \at{Numerical simulation of particle-laden turbulent channel
  flow}.  \jt{Phys. Fluids}  \bvol{13}~(10),  \pg{2957--2967}.

\bibitem[Majji {\em et~al.\/}(2018)Majji, Banerjee \&
  Morris]{majji2018inertial}
{\sc \au{Majji, M.~V.}, \au{Banerjee, S.} \& \au{Morris, J.~F.}} \yr{2018}
  \at{Inertial flow transitions of a suspension in taylor--couette geometry}.
  \jt{J. Fluid Mech.}  \bvol{835},  \pg{936--969}.

\bibitem[Massot(2007)]{massot2007eulerian}
{\sc \au{Massot, M.}} \yr{2007}  \at{Eulerian multi-fluid models for
  polydisperse evaporating sprays}.  \bt{In {\em Multiphase reacting flows:
  modelling and simulation\/}},  \pg{pp. 79--123}.  \publ{Springer}.

\bibitem[Matas {\em et~al.\/}(2003)Matas, Morris \&
  Guazzelli]{matas2003transition}
{\sc \au{Matas, J.-P.}, \au{Morris, J.~F.} \& \au{Guazzelli, E.}} \yr{2003}
  \at{Transition to turbulence in particulate pipe flow}.  \jt{Phys. Rev.
  Lett.}  \bvol{90},  \pg{014501}.

\bibitem[Maxey(1987)]{maxey1987motion}
{\sc \au{Maxey, M.}} \yr{1987}  \at{The motion of small spherical particles in
  a cellular flow field}.  \jt{The Physics of fluids}  \bvol{30}~(7),
  \pg{1915--1928}.

\bibitem[Maxey \& Riley(1983)]{maxey1983equation}
{\sc \au{Maxey, M.~R.} \& \au{Riley, J.~J.}} \yr{1983}  \at{Equation of motion
  for a small rigid sphere in a nonuniform flow}.  \jt{Phys. Fluids}
  \bvol{26}~(4),  \pg{883--889}.

\bibitem[Michael(1964)]{michael1964stability}
{\sc \au{Michael, D.}} \yr{1964}  \at{The stability of plane poiseuille flow of
  a dusty gas}.  \jt{J. Fluid Mech.}  \bvol{18}~(1),  \pg{19--32}.

\bibitem[Pan \& Banerjee(1996)]{Pan1996PoF}
{\sc \au{Pan, Y.} \& \au{Banerjee, S.}} \yr{1996}  \at{Numerical simulation of
  particle interactions with wall turbulence}.  \jt{Phys. Fluids}
  \bvol{8}~(10),  \pg{2733--2755}.

\bibitem[Pope(2000)]{pope2000turbulent}
{\sc \au{Pope, S.~B.}} \yr{2000} {\em Turbulent flows\/}.  \publ{Cambridge
  university press}.

\bibitem[Richter(2015)]{richter2015turbulence}
{\sc \au{Richter, D.~H.}} \yr{2015}  \at{Turbulence modification by inertial
  particles and its influence on the spectral energy budget in planar couette
  flow}.  \jt{Phys. Fluids}  \bvol{27}~(6),  \pg{063304}.

\bibitem[Richter \& Sullivan(2013)]{richter2013momentum}
{\sc \au{Richter, D.~H.} \& \au{Sullivan, P.~P.}} \yr{2013}  \at{Momentum
  transfer in a turbulent, particle-laden couette flow}.  \jt{Phys. Fluids}
  \bvol{25}~(5),  \pg{053304}.

\bibitem[Richter \& Sullivan(2014)]{richter2014modification}
{\sc \au{Richter, D.~H.} \& \au{Sullivan, P.~P.}} \yr{2014}  \at{Modification
  of near-wall coherent structures by inertial particles}.  \jt{Phys. Fluids}
  \bvol{26}~(10),  \pg{103304}.

\bibitem[Rudyak {\em et~al.\/}(1998)Rudyak, Isakov \&
  Bord]{rudyak1998instability}
{\sc \au{Rudyak, V.~Y.}, \au{Isakov, E.} \& \au{Bord, E.}} \yr{1998}
  \at{Instability of plane couette flow of two-phase liquids}.  \jt{Tech. Phys.
  Lett.}  \bvol{24}~(3),  \pg{199--200}.

\bibitem[Rudyak {\em et~al.\/}(1997)Rudyak, Isakov \&
  Bord]{rudyak1997hydrodynamic}
{\sc \au{Rudyak, V.~Y.}, \au{Isakov, E.~B.} \& \au{Bord, E.~G.}} \yr{1997}
  \at{Hydrodynamic stability of the poiseuille flow of dispersed fluid}.
  \jt{J. Aerosol Sci.}  \bvol{28}~(1),  \pg{53--66}.

\bibitem[Saffman(1962)]{saffman1962stability}
{\sc \au{Saffman, P.}} \yr{1962}  \at{On the stability of laminar flow of a
  dusty gas}.  \jt{J. Fluid Mech.}  \bvol{13}~(1),  \pg{120--128}.

\bibitem[Sardina {\em et~al.\/}(2012)Sardina, Schlatter, Brandt, Picano \&
  Casciola]{sardina2012wall}
{\sc \au{Sardina, G.}, \au{Schlatter, P.}, \au{Brandt, L.}, \au{Picano, F.} \&
  \au{Casciola, C.~M.}} \yr{2012}  \at{Wall accumulation and spatial
  localization in particle-laden wall flows}.  \jt{J. Fluid Mech.}  \bvol{699},
   \pg{50--78}.

\bibitem[Schiller(1933)]{schiller1933ber}
{\sc \au{Schiller, V.}} \yr{1933}  \at{ber die grundlegenden berechnungen bei
  der schwerkraftaufbereitung}.  \jt{Z. Vereines Deutscher Inge.}  \bvol{77},
  \pg{318--321}.

\bibitem[Schoppa \& Hussain(2002)]{schoppa2002coherent}
{\sc \au{Schoppa, W.} \& \au{Hussain, F.}} \yr{2002}  \at{Coherent structure
  generation in near-wall turbulence}.  \jt{J. Fluid Mech.}  \bvol{453},
  \pg{57--108}.

\bibitem[Sendstad \& Moin(1992)]{sendstad1992near}
{\sc \au{Sendstad, O.} \& \au{Moin, P.}} \yr{1992} The near-wall mechanics of
  three-dimensional boundary layers. rep. tf-57. thermoscience divison,
  department of mechanical engineering.

\bibitem[Smits {\em et~al.\/}(2011)Smits, McKeon \& Marusic]{smits2011high}
{\sc \au{Smits, A.~J.}, \au{McKeon, B.~J.} \& \au{Marusic, I.}} \yr{2011}
  \at{High--reynolds number wall turbulence}.  \jt{Annu. Rev. Fluid Mech.}
  \bvol{43}.

\bibitem[Squires \& Eaton(1990)]{squires1990particle}
{\sc \au{Squires, K.~D.} \& \au{Eaton, J.~K.}} \yr{1990}  \at{Particle response
  and turbulence modification in isotropic turbulence}.  \jt{Phys. Fluids A}
  \bvol{2}~(7),  \pg{1191--1203}.

\bibitem[Stone {\em et~al.\/}(2004)Stone, Roy, Larson, Waleffe \&
  Graham]{stone2004polymer}
{\sc \au{Stone, P.~A.}, \au{Roy, A.}, \au{Larson, R.~G.}, \au{Waleffe, F.} \&
  \au{Graham, M.~D.}} \yr{2004}  \at{Polymer drag reduction in exact coherent
  structures of plane shear flow}.  \jt{Phys. Fluids}  \bvol{16}~(9),
  \pg{3470--3482}.

\bibitem[Sweet {\em et~al.\/}(2018)Sweet, Richter \& Thain]{sweet2018gpu}
{\sc \au{Sweet, J.}, \au{Richter, D.~H.} \& \au{Thain, D.}} \yr{2018}  \at{Gpu
  acceleration of eulerian--lagrangian particle-laden turbulent flow
  simulations}.  \jt{Int. J. Multiphase Flow}  \bvol{99},  \pg{437--445}.

\bibitem[Tanaka \& Eaton(2008)]{tanaka2008classification}
{\sc \au{Tanaka, T.} \& \au{Eaton, J.~K.}} \yr{2008}  \at{Classification of
  turbulence modification by dispersed spheres using a novel dimensionless
  number}.  \jt{Phys. Rev. Lett.}  \bvol{101}~(11),  \pg{114502}.

\bibitem[Waleffe(1997)]{waleffe1997self}
{\sc \au{Waleffe, F.}} \yr{1997}  \at{On a self-sustaining process in shear
  flows}.  \jt{Phys. Fluids}  \bvol{9}~(4),  \pg{883--900}.

\bibitem[Wang {\em et~al.\/}(2017)Wang, Abbas \& Climent]{wang2017modulation}
{\sc \au{Wang, G.}, \au{Abbas, M.} \& \au{Climent, E.}} \yr{2017}
  \at{Modulation of large-scale structures by neutrally buoyant and inertial
  finite-size particles in turbulent couette flow}.  \jt{Phys. Rev. Fluids}
  \bvol{2}~(8),  \pg{084302}.

\bibitem[Wang {\em et~al.\/}(2018)Wang, Abbas \& Climent]{wang2018JFM}
{\sc \au{Wang, G.}, \au{Abbas, M.} \& \au{Climent, E.}} \yr{2018}
  \at{Modulation of the regeneration cycle by neutrally buoyant finite-size
  particles}.  \jt{accepted by J. Fluid Mech., arXiv preprint arXiv:1806.02862}
  .

\bibitem[Yu {\em et~al.\/}(2016)Yu, Vinkovic \& Buffat]{yu2016finite}
{\sc \au{Yu, W.}, \au{Vinkovic, I.} \& \au{Buffat, M.}} \yr{2016}
  \at{Finite-size particles in turbulent channel flow: quadrant analysis and
  acceleration statistics}.  \jt{J Turbul}  \bvol{17}~(11),  \pg{1048--1071}.

\bibitem[Zhao {\em et~al.\/}(2013)Zhao, Andersson \&
  Gillissen]{zhao2013interphasial}
{\sc \au{Zhao, L.}, \au{Andersson, H.~I.} \& \au{Gillissen, J.~J.}} \yr{2013}
  \at{Interphasial energy transfer and particle dissipation in particle-laden
  wall turbulence}.  \jt{J. Fluid Mech.}  \bvol{715},  \pg{32}.

\end{thebibliography}

\end{document}